\documentclass{aa}

\usepackage{graphicx}
\usepackage{txfonts}
\usepackage{hyperref}
\usepackage{xcolor}
\hypersetup{
  colorlinks   = true, 
  urlcolor     = black, 
  linkcolor    = blue, 
  citecolor    = blue 
}
\usepackage[all]{hypcap} 
\usepackage{mathtools}

\begin{document}

    \title{Probing the kinematics of the Local Group with chemically enriched gas in the {\sc Hestia} simulations}
    \titlerunning{Kinematics and chemical abundances in the Local Group}

    \author{
L. Biaus\inst{1}\thanks{E-mail: lbiaus@df.uba.ar} \and 
S.~E.~Nuza\inst{2} \and 
C.~Scannapieco\inst{1} \and 
P.~Richter\inst{3, 4} \and 
M.~Damle \inst{5, 6} \and
N.~I.~Libeskind \inst{4} \and
M.~Vogelsberger \inst{7}
}

    \institute{
Universidad de Buenos Aires, Facultad de Ciencias Exactas y Naturales, Departamento de Física, Buenos Aires, Argentina \and
Instituto de Astronomía y Física del Espacio (IAFE, CONICET-UBA), CC 67, Suc. 28, 1428 Buenos Aires, Argentina \and
Institut f{\"u}r Physik und Astronomie, Universität Potsdam, Karl-Liebknecht-Str. 24/25, 14476 Golm, Germany \and
Leibniz-Institut f{\"u}r Astrophysik, An der Sternwarte 16, 14482 Potsdam, Germany \and
New York University Abu Dhabi, Department of Physics, PO Box 129188, Abu Dhabi, UAE \and
Center for Astrophysics and Space Science (CASS), New York University Abu Dhabi, PO Box 129188, Abu Dhabi, UAE \and
Department of Physics, Kavli Institute for Astrophysics and Space Research, Massachusetts Institute of Technology, Cambridge, MA 02139, USA
}

%   \date{Received September 15, 1996; accepted March 16, 1997}
	\date{}

\abstract{We present a study of the gas kinematics within the {\sc Hestia} project, a state-of-the-art set of simulations of the Local Group, with a particular focus on the velocity patterns of different ions and the large-scale motion of gas and galaxies towards the Local Group's barycentre. Using two of the {\sc Hestia} high-resolution runs, we examined the distribution and velocities of H\,{\sc i}, C\,{\sc iv}, Si\,{\sc iii}, O\,{\sc vi}, O\,{\sc vii}, and O\,{\sc viii} and their imprints on sightlines observed from the Sun's location in different reference frames. To mimic observational strategies, we assessed the contribution of rotating disc gas, assuming simple kinematic and geometrical considerations. Our results indicate that local absorption features in observed sightlines most likely trace material in the circumgalactic medium of the Milky Way. Some sightlines, however, show that intragroup material could be more easily observed towards the barycentre, which defines a preferred direction in the sky. In particular, H\,{\sc i}, Si\,{\sc iii}, and C\,{\sc iv} roughly trace cold gas inside Milky Way and Andromeda haloes, as most of their mass flux occurs within the virial region of each galaxy, while oxygen high ions mostly trace hot halo and intragroup gas, with comparable mass fluxes in the Local Group outskirts and the circumgalactic medium of the two main galaxies. Additionally, we find that pressures traced by different ionic species outside the Milky Way's halo show systematically higher values towards the barycentre direction in contrast to its antipode in the sky. Kinematic imprints of the global motion towards the barycentre can be seen at larger distances for all ionic species as the Milky Way rams into material in the direction of Andromeda, with gas towards the anti-barycentre lagging behind.}


\maketitle

\section{Introduction}

Galaxies in our close cosmological vicinity are part of the so-called Local Group (LG), a loose aggregation of galactic systems dominated by the Milky Way (MW) and its neighbour Andromeda (M31), thus comprising most of the group's mass. The global LG kinematics is, therefore, essentially driven by the relative motion between MW, M31, and their satellite galaxies. As a result, galaxies in the LG are part of a general flow towards the LG barycentre and it is believed that MW and M31 will experience their first core passage in the next few gigayears \citep[e.g.][]{Salomon21}.

\begin{figure*}
    \centering
    \includegraphics[width=1.7\columnwidth]{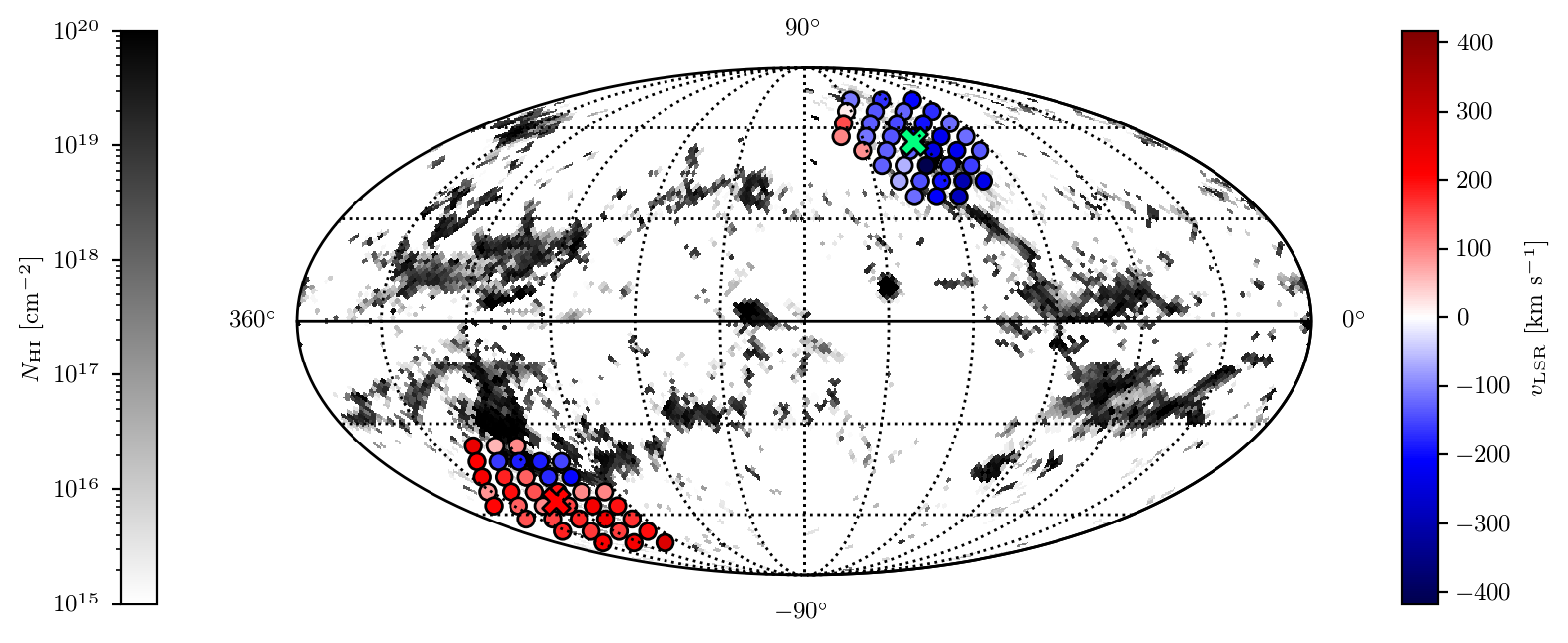}\caption{
    Column density map of H\,{\sc i} high-velocity clouds in simulation $17\_11$, as seen by a simulated observer from the Sun's position. Colour-coded LSR gas velocities in sightlines evenly spread towards the general barycentre (green cross) and anti-barycentre (red cross) directions are also shown. Galactic longitude increases from right to left.
    }
    \label{fig:fig1}
\end{figure*}

The radial motion between MW and M31 determines a preferred direction along which the LG kinematics should be more evident. In fact, observations indicate that LG galaxies display a velocity dipole along the general direction connecting the M31 region to its antipode in the sky at relatively low and high galactic latitudes \citep[see e.g.][and references therein]{Richter17}. This suggests that galaxies located in the general anti-barycentre direction are lagging behind the MW flow speed, while galaxies approaching the LG barycentre from the opposite side are moving towards the MW. Along these lines, \cite{Richter17} propose a simple picture in which all gas in the LG -- including gas in the intragroup medium (IGrM) -- shares the same global motion towards the group's barycentre. This scenario would naturally explain the origin of a similar velocity dipole observed for a sample of high-velocity absorbers that originated from diffuse gas in our cosmic vicinity if some of them were located outside the circumgalactic medium (CGM) of the MW, i.e. far from the, presumably, more turbulent Galactic halo. In particular, \cite{Bouma19} modelled the gas ionisation conditions of low and intermediate ions and conclude that a non-negligible fraction of high-velocity clouds lying close to the LG barycentre have low thermal gas pressure values that are consistent with the properties of ambient gas in the IGrM, i.e. possibly located outside the virial radius of the MW.

In \cite{Biaus22}, the previous paper in this series, we focus on the kinematics of the LG and its imprint on the gas and galaxy velocity maps as seen from an observer located at the Sun's position using the three high-resolution LG simulations of the {\sc Hestia} project \citep{Libeskind20, Dupuy2022, Khoperskov2023c, Khoperskov2023a, Khoperskov2023b}. These simulations are state-of-the-art numerical representations of LG-like regions sharing several important characteristics with the actual LG, such as the presence of two interacting MW-mass haloes similar to the MW-M31 system embedded in a filament surrounded by the most relevant large-scale structures in the local Universe.

The existence of a wide, large-scale radial velocity dipole in the sky resembling observations from the Local Standard of Rest (LSR) is a natural outcome of these simulations. This is expected as the rotation of the MW-like galaxy candidates imprints a strong asymmetry in the sky, especially along the direction of the Sun's velocity vector. For sightlines above the galactic plane, rotation plays a minor role and the resulting velocity maps may reflect the relative motion between MW and M31, which is more clearly seen from the so-called Galactic Standard of Rest (GSR) reference frame. 

In particular, we have shown that when the MW and M31 candidates are, in fact, radially approaching, a significant fraction of gas outside the MW's virial radius located in front of and behind the Sun's position shows negative and positive velocities, respectively, as seen from the GSR. This result is consistent with the interpretation given by \cite{Richter17} for at least some fraction of the absorbing material, where the observer in the MW moves faster than material lagging behind, while it rams into LG gas that is at rest in the barycentre frame.

 In this work, we aim to analyse the contribution of six different ionic species to the dipole pattern expected for IGrM\footnote{Throughout this work, the intragroup medium (IGrM) refers to all gas located beyond $R_{200}$ of both the MW and M31 galaxies without any regard to energy considerations.} gas. This was achieved by analysing how these ionic species are spatially distributed (e.g. within or outside the MW-like galactic haloes) and taking into account their respective kinematics.

 This paper is organised as follows. In Section 2, we briefly present the main aspects of the LG simulations analysed here. In Section 3, we describe the setup for the velocity reference frames implemented at the Sun’s position in the simulated MWs. We also show the spherically averaged gas mass profiles within the LG, and beyond, including and excluding virialised material. In Section 4, we present the predictions for the kinematics of gas and galaxies in the simulations. Finally, in Section 5, we summarise this work and discuss our results in light of current observations.

\section{Simulations}

Following \cite{Biaus22}, we use the {\sc Hestia} simulations to reproduce galaxy systems which resemble the LG within a cosmological context. In this section, we present a summary of the main characteristics of the three high-resolution runs ($37\_11$, $9\_18$ and $17\_11$) in the {\sc Hestia} project. More details concerning the cosmological code, the galaxy formation model and the different LG realisations can be found in \cite{Libeskind20} and references therein.

\subsection{Simulation code}

{\sc Hestia} simulations were run using the moving-mesh {\sc Arepo} cosmological code \citep{Springel10, Weinberger20}. {\sc Arepo} solves the gravitational forces and ideal magnetohydrodynamics (MHD) to compute the joint evolution of gas, stars and dark matter (DM). The equations are coupled to the {\sc Auriga} galaxy formation model \citep{Pakmor13, Grand17}. The code generates a Voronoi mesh from a set of points which move with the gas velocity and minimises advection errors when compared to static mesh codes.

The main physical processes implemented within the {\sc Auriga} galaxy formation model are primordial and metal-dependent gas cooling, a redshift-dependent ultraviolet (UV) background \citep{FaucherGiguere09}, star formation using a Chabrier initial mass function \citep{SpringelHernquist03, Chabrier03}, chemical and energy feedback from core-collapse and Type Ia supernovae, asymptotic giant branch stars and active galactic nuclei coupled to the formation and evolution of supermassive black holes. We refer the reader to the works of \cite{Grand17} and \cite{Weinberger20} for further details. For a review on cosmological simulations of galaxy formation, we refer the reader to \cite{Vogelsberger2020}.

\subsection{Local Group simulations}

The cosmological parameters assumed in the simulations used in this work are those derived from the best fit of \cite{Planck14}: $\Omega_{\Lambda}=0.682$, $\Omega_{\rm M}=0.270$, $\Omega_{\rm b}=0.048$, $\sigma_8=0.83$ and $H_0=100\,h$\,km\,s$^{-1}$\,Mpc$^{-1}$ with $h=0.677$.
Observational data from our cosmic neighbourhood is taken from the CosmicFlows-2 catalogue \citep{Tully13} to construct the initial conditions (ICs) in a way such that they recreate the local Universe's large-scale structure and peculiar velocities at $z=0$. Around a thousand of DM-only, low-resolution ICs are generated from these constraints, within periodic boxes of $100\,h^{-1}\,$Mpc on a side, filled with $256^3$ DM particles. A zoom-in technique is used to populate a central region with $512^3$ effective particles of mass $m_{\rm DM} = 6 \times 10^8$~M$_\odot$, which replace their counterparts in the low-resolution run. This region encompasses a sphere of radius $10\,h^{-1}\,$Mpc. The resulting number of particles ensures that there will be a few thousand particles conforming each of the two main LG haloes.

The simulations aim to reproduce the main cosmographic features of the LG, namely: the Virgo cluster, the local void and the local filament. An algorithm scans the set of ICs for LG-like regions resembling the actual MW-M31 system containing halo pairs with mass ratios no greater than 2. In each realisation, the more massive halo is termed as `M31' and the other major halo as `MW'. Having selected potential LG-like regions, the ICs are regenerated at a higher resolution, using $4096^3$ effective particles within a sphere of radius  $5\,h^{-1}\,$Mpc. These particles are subsequently split into a DM-gas cell pair respecting the cosmic baryon fraction. Three ICs from this sample were rerun at an even higher resolution from which only two (namely, realisations $9\_18$ and $17\_11$) are analysed here\footnote{LG realisation $37\_11$ was not analysed in this work because the relative infall velocity of the two main haloes at $z=0$ ($v_{\rm rad}\approx 9\,$km\,s$^{-1}$) is at odds with the observed value of $-109 \pm 4.4\,$km\,s$^{-1}$ \citep{vanderMarel12} as it was shown in \cite{Biaus22}.}.  In these higher resolution runs, two overlapping spheres of diameter $2.5\,h^{-1}\,$Mpc are drawn around the two main LG halos, populated with $8192^3$ effective particles. This results in particle masses of $m_{\rm DM} = 1.5 \times 10^5$~M$_\odot$ for the dark matter, $m_{\rm gas} = 2.2 \times 10^4$~M$_\odot$ for the gas, and a spatial resolution of $\epsilon = 220$~pc. In realisations $9\_18$ and $17\_11$, the two main haloes are approaching each other at an infall radial velocity of $-74$ and $-102\,$km\,s$^{-1}$, respectively. The main characteristics of the MW and M31 candidates in both simulations are summarised in Table \ref{table:1}.

\begin{table}
\caption{MW and M31 properties.}
\label{table:1}
\centering
\begin{tabular}{c c c c c} 
 \hline\hline
 {LG realisation}  & \multicolumn{2}{c}{$9\_18$} & \multicolumn{2}{c}{$17\_11$} \\ 

 {Galaxy} & MW & M31 & MW & M31\\

 \hline\
 $R_{200}$ (kpc) & 254 & 262 & 255 & 269\\ 
 $M_{200}$ ($10^{12}\,$M$_{\odot}$) & 1.88 & 2.06 & 1.89 & 2.23\\ 
 $M_{\rm gas}$ ($10^{10}\,$M$_{\odot}$) & 15.1 & 14.7 & 10.5 & 16.3\\
 $M_{\star}$ ($10^{10}\,$M$_{\odot}$) & 10.8 & 12.8 & 11.4 & 12.6\\
  $f_{gas}$ & 0.08 & 0.071 & 0.056 & 0.073\\
 $f_{\star}$ & 0.057 & 0.062 & 0.06 & 0.057\\
 $v_{\rm rad}$ (km s$^{-1}$) & \multicolumn{2}{c}{$-74$} & \multicolumn{2}{c}{$-102$}\\
 $d$ (kpc) & \multicolumn{2}{c}{$866$} & \multicolumn{2}{c}{$675$}\\ 

 \hline
 $r_{\rm max}$ (kpc) & \multicolumn{2}{c}{\bf $50$} & \multicolumn{2}{c}{$50$}\\ 
 $z_{\rm max}$ (kpc) & \multicolumn{2}{c}{\bf $8$} & \multicolumn{2}{c}{$8$}\\
 $v_{\rm dev}$ (km s$^{-1}$) & \multicolumn{2}{c}{$100$} & \multicolumn{2}{c}{$100$}\\%[1ex] 
 \hline\hline
 \vspace{.01cm}
\end{tabular}
\tablefoot{Main properties of the MW and M31 galaxy candidates for the two {\sc Hestia} high-resolution runs studied in this work. From top to bottom: the distance where the mass density profile of haloes equals 200 times the critical density of the Universe, $R_{200}$; the total, gas and stellar masses within $R_{200}$; the relative radial velocity, $v_{\rm rad}$, and distance, $d$, between MW and M31. The last three rows show the parameters chosen to filter out the gaseous discs of the galaxy candidates using the method presented in Section~\ref{sec:filtering}.}
\end{table}

In what follows, we primarily focus on the high-resolution realisation $17\_11$ which, as discussed in \cite{Biaus22}, is our best LG-like candidate in terms of global kinematics. This does not qualitatively change our results since both LG simulations share the same global trends. However, to gauge expected differences in the galactic halo properties of the MW candidate, we occasionally refer to realisation $9\_18$ in Sections~\ref{sec:chem_kin} and~\ref{sec:chem_pressure}.

\section{Analysis} 
\label{sec:analysis}

\subsection{Reference frames}
\label{sec:reference}

Throughout this paper, we refer to three widely used frames of reference usually adopted by observers to measure line-of-sight velocities. These are the LSR, the GSR, and the LG Standard of Rest (LGSR). In the LSR, velocities are referred to the mean velocity of the Sun's immediate vicinity. In the GSR, the tangential velocity component of the Sun around the MW is removed, so velocities are referred to the MW's bulk velocity. In the LGSR, the radial motion of the MW with respect to the LG barycentre is additionally excluded \citep[e.g.][]{Karachentsev96, Courteau99}.

To define our galactic coordinate system in the simulations, we compute the angular momentum of the MW stellar disc and take the north Galactic pole in the opposite direction, as required by the usual definition of Galactic coordinates. We position the Sun in the midplane of our simulated MW at $r = 8\,$kpc from its centre and at an azimuthal position from which the galactic longitude of the simulated M31 matches the observed one. The synthetic LSR is defined assuming that the velocity of the Sun is given by a purely tangential motion with an absolute velocity value determined by the rotation curve of the MW analogue at the Sun's location in each simulation. The GSR velocities are obtained by removing this tangential velocity, and the LGSR velocities can be obtained from GSR velocities by further removing the radial motion of the MW with respect to the LG barycentre. For further details on how we build these reference frames we refer the reader to \cite{Biaus22}.

\begin{figure*}   
 \centering
 \sidecaption
 \includegraphics[width=10cm]{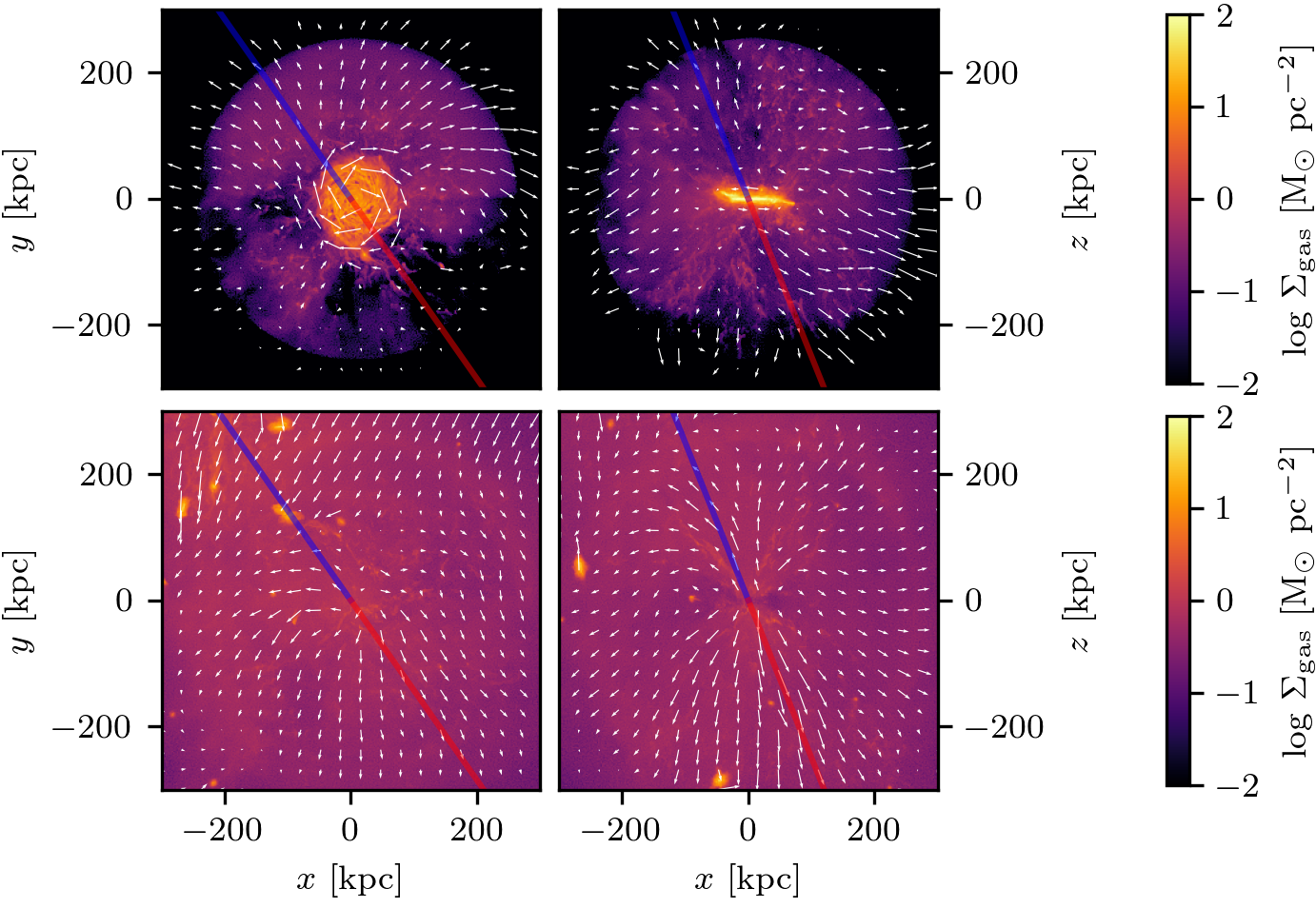}
 \vspace{.5cm}
    \caption{Gas distribution at $z=0$ obtained after applying the kinematic filtering method described in Section~\ref{sec:filtering} for the MW candidate in LG simulation $17\_11$ for face-on (left panels) and edge-on (right panels) views. The upper (lower) panels show material consistent (inconsistent) with disc-like velocities according to the adopted filter. The velocity field of gas in all panels is indicated by the white arrows. The blue and red solid lines in all panels indicate the barycentre and anti-barycentre directions, respectively.}
    \label{fig:map_filter}
\end{figure*}

In Fig.~\ref{fig:fig1} we present a sky map of {\sc H\,i} high-velocity clouds (HVCs) in simulation $17\_11$, as seen from the LSR reference frame. Additionally, the mass-weighted average gas velocities in the LSR are shown for different sightlines pointing towards the barycentre and anti-barycentre directions within an opening angle of $20^{\circ}$ (solid circles). This angular extent is motivated by the observational study of \cite{Bouma19} with emphasis along the LG barycentre region. These authors analysed a sample of absorbers selected to characterise gas along the barycentre/anti-barycentre direction. In agreement with these observations, a dipolar trend between these two general directions is seen, where negative (positive) velocities dominate the barycentre (anti-barycentre) regions thus suggesting a connection of these absorbers with the global motion between MW and M31, as already shown in \cite{Biaus22}. Throughout this work, we adopt line-of-sight cones of $20^{\circ}$ to characterise the kinematic, chemical and thermodynamical properties of gas along the preferred infall direction of the LG. 

\subsection{Filtering out the gaseous disc}
\label{sec:filtering}

To build the neutral HVC map shown in Fig.~\ref{fig:fig1}, we used a mask to filter out disc gaseous material using the method described in \cite{Westmeier18}. In that work, the authors modelled the rotation of the MW disc assuming cylindrical symmetry with velocity vectors parallel to the Galactic plane in a purely tangential direction and a magnitude given by the rotation curve of the Galaxy, independently of the height above the disc. With this cylindrical model in mind, the authors calculated the radial velocity of an arbitrary point in the disc as seen from the Sun, arriving at the following expression

\begin{equation}
    v_{\rm rad} = \left( v_{\rm rot}(r_{xy}) \frac{r_{\odot}}{r_{xy}} - v_{\odot} \right) \sin(l)\cos(b),
    \label{vrad}
\end{equation}

\noindent where $v_{\rm rot}(r_{xy})$ is the rotational velocity of the MW disc at distance $r_{xy}$ from the Galactic centre\footnote{The radial distance $r_{xy}$ is obtained after projecting the position vector $\mathbf{r} = (r_x, r_y, r_z)$ of a point in the plane of the Galactic disc.}, $r_{\odot}$ is the Sun's distance to the Galactic centre, $v_{\odot}$ is the Sun's rotational velocity and $(l, b)$ are the Galactic longitude and latitude, respectively. We note that, when using Galactic coordinates, the $x$--axis points in the direction of the Sun's rotation, the $y$--axis points away from the Galactic centre and the $z$--axis points in the direction of the north Galactic pole.

Using Eq.~\ref{vrad}, the authors established the radial velocity span of the Galactic disc gas at all positions in the sky. This was achieved by incrementally moving away from the Sun along the line of sight, assessing the radial velocity of the gas at each step, and calculating the minimum ($v_{\rm min}$) and maximum ($v_{\rm max}$) values found between the starting point and the edge of the cylindrical disc model, which is determined by the disc size parameters, $r_{\rm max}$ and $z_{\rm max}$. Finally, to account for the velocity dispersion in the disc gas, the derived velocity ranges are expanded by a fixed deviation velocity, $v_{\rm dev}$. This leads to a final velocity range, i.e. $[v_{\rm min} - v_{\rm dev}, v_{\rm max} + v_{\rm dev}$], that is used to mask the radial velocity data, leaving only channels inconsistent with Galactic rotation according to the adopted disc rotation model. For more details on this procedure, we refer the reader to the original publication.

Throughout this work, we followed the \cite{Westmeier18} method to mask out material consistent with a coherent motion of the gaseous disc, which simplifies the comparison with observational results. To account for the galactic rotation curve in the simulations, we computed the corresponding circular velocity at distance $r$ from the galactic centre as

\begin{equation}
    v_{\rm rot}(r) = \sqrt{\frac{G M(r)}{r}},
\end{equation}

\noindent where $M(r)$ is the total mass enclosed at radius $r$ from the simulated galaxy centre. Parameters $r_{\rm max}$, $z_{\rm max}$ and $v_{\rm dev}$ were chosen within reasonable limits for each of the model galaxies by visual inspection of the resulting HVC maps, and are reported in Table~\ref{table:1} for the two LG simulations analysed here.

The upper panels of Fig.~\ref{fig:map_filter} show the excluded gaseous material consistent with disc rotation after applying the filter described above to the gas distribution within $R_{200}$ for the MW candidate in simulation $17\_11$ for both face-on (left column) and edge-on (right column) views. Conversely, the lower panels show all gas that remains within the MW halo after the filter is applied. In all panels, the blue (red) lines indicate the barycentre (anti-barycentre) directions, with the blue lines roughly indicating the direction of the M31 candidate within the LG volume. As expected, the upper panels clearly show that the filtering method is able to identify most of the rotating material belonging to the gaseous disc and its neighbourhood within the galactic halo, thus allowing us to identify material with different kinematic properties. We note, however, that the resulting kinematically segregated gas distribution comprises all gas phases so that any further classification of material surrounding the galaxy, such as the non-rotating synthetic HVCs shown in Fig.~\ref{fig:fig1} for instance, must consider other selection criteria such as temperature, composition and/or ionisation level.

\section{Results}
\label{sec:results}

\subsection{Chemically enriched gas distribution}

\begin{figure*}
    \centering
    \includegraphics[width=1.6\columnwidth]{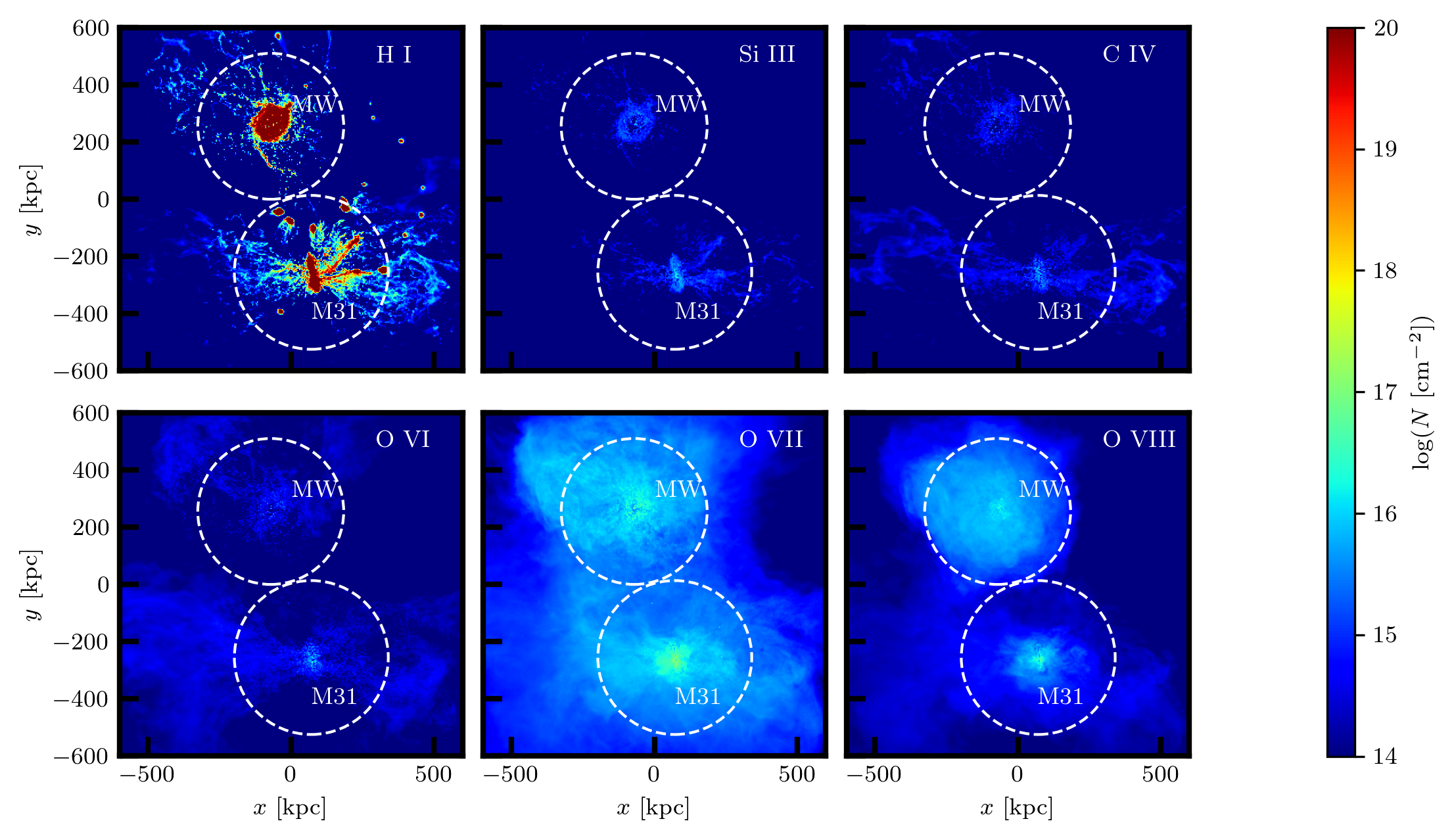}
    \caption{Distribution of the six studied ions at $z=0$ in the simulated Local Group realisation $17\_11$. Dashed circles indicate the virial radius of each galactic system. The colourbar indicates the column density of each ionic species.%\Sebas{Corregir los $R_{200}$ y marcarlos en el plot.} 
    }
    \label{fig:LG_ion_distribution}
\end{figure*}

The chemical abundances of gas within the LG simulations are computed using ionisation tables for the {\sc H\,i}, {\sc C\,iv}, {\sc Si\,iii}, {\sc O\,vi}, {\sc O\,vii} and {\sc O\,viii} ionic species calculated by \cite{Hani18} for the {\sc Hestia} high-resolution volumes. These tables were computed using the spectral synthesis code {\sc Cloudy} under the assumption of chemical ionisation equilibrium \citep{Ferland17}. We note that departures from equilibrium can be relevant in the CGM and IGrM. In particular, non-equilibrium effects may arise when the gas is rapidly cooling or being shock-heated, which can alter the relative ion abundances \citep[e.g.][]{Gnat2007,Oppenheimer2013}. In addition, our tables assume a uniform UV background, neglecting the contribution of ionising photons from stars within the MW and M31, which could also alter ionisation fractions in the inner CGM. However, a fully self-consistent treatment would require modelling time-dependent ionisation and local radiation fields, which is beyond the scope of this work.

In this work, we want to characterise gas properties along the preferred barycentre and anti-barycentre directions, as done in \cite{Biaus22}, but focusing on the different chemical element distributions. In Fig. \ref{fig:LG_ion_distribution}, we show the distribution of ions in the LG realisation $17\_11$. As seen in the figure, the neutral H\,{\sc i} component reaches very high column densities in the inner parts of galaxies where most of the cold, dense gas is located. This behaviour is also seen in the less-abundant C\,{\sc iv} and Si\,{\sc iii} ions, although they are also to be found in more filamentary structures located in the IGrM that are linked to galactic outflows. 

In general, all ions but H\,{\sc i} display similar column densities in the inner and outer galactic regions, with O\,{\sc vii} and O\,{\sc viii} having the largest difference between the ISM of the galaxies and the IGrM. In both systems, O\,{\sc vii} tend to trace their large-scale, circumgalactic coronae, whereas the highest O\,{\sc viii} ions trace the hotter gas components. In fact, in this particular realisation, O\,{\sc viii} extends further out from the MW as this galaxy presents strong, hot outflows at $z=0$ which ionise the CGM up to a distance of about $R_{200}$ \citep{Biaus22,Damle22}. We have verified that, despite the presence of strong outflows, the hot CGM remains close to hydrostatic equilibrium, although deviations occur at several radii (see Fig. \ref{fig:HSE} in Appendix \ref{app:HSE}).

\subsubsection{Line-of-sight column densities}

\begin{figure}
    \centering
    \includegraphics[width=.9\columnwidth]{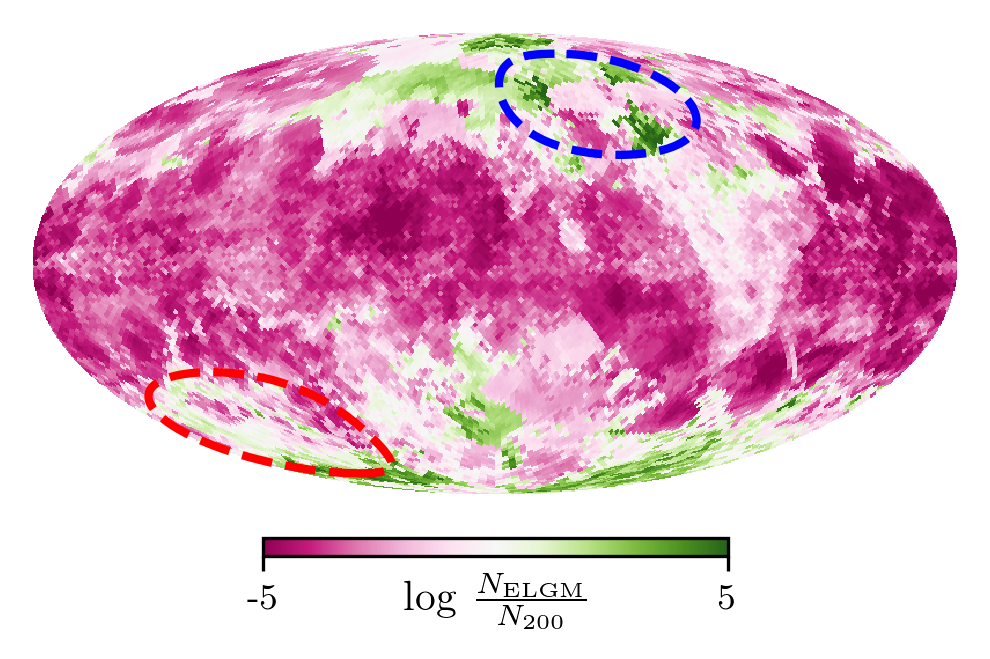}
    \caption{Ratio of the ELGM column density contribution to the CGM contribution for C\,{\sc iv}. Directions showing green regions have a higher contribution from the IGrM than from the galactic halo. Blue (red) dashed lines indicate barycentre (anti-barycentre) regions in the sky for an opening angle of $20^{\circ}$. 
    }
    \label{fig:CIV_ratio_mollweide}
\end{figure}

\begin{figure*}
    \centering
    \includegraphics[width=1.8\columnwidth]{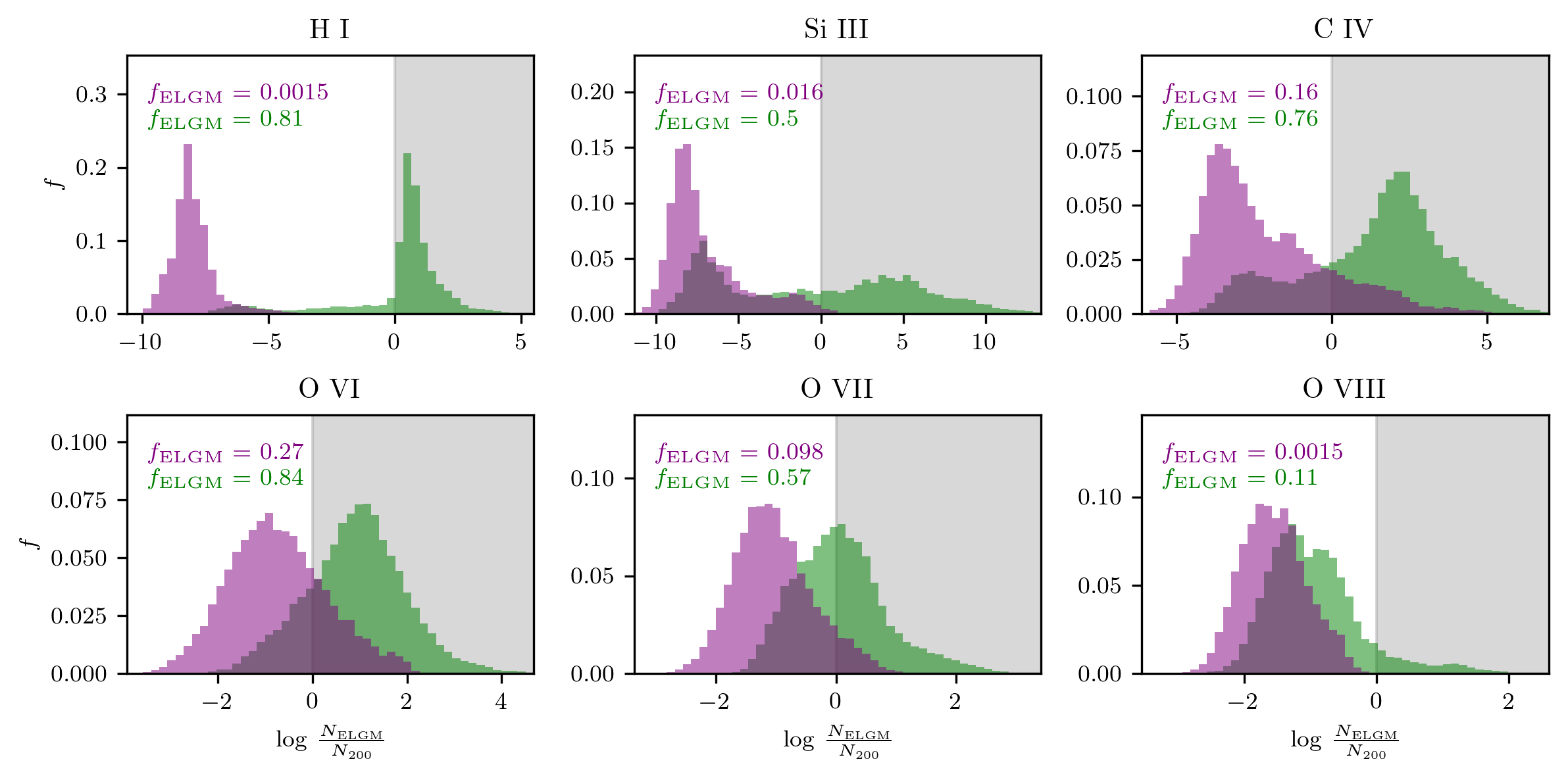}
    \caption{Distribution of column density ratio $N_{\rm ELGM}/N_{200}$ for the six studied ions, for a sample of sightlines covering the simulated sky of simulation $17\_11$ uniformly. Purple histograms correspond to distributions for all material within and outside the CGM of the MW candidate, while green histograms show the distributions after material with disc-like motion within $R_{200}$ has been filtered out using the method described in Sec.~\ref{sec:filtering}. The fraction of sightlines with column density ratio $N_{\rm ELGM}/N_{200} > 1$ (grey shaded area), noted as $f_{\rm ELGM}$, is indicated in all panels for each histogram.
    }
    \label{fig:N_ratio_hist}
\end{figure*}

The expected higher column densities associated with the disc and the CGM compared to those of the IGrM raise the question of whether certain ions are more likely to give us information about the IGrM than others. To address this question, we compared the column density contributions from external material with those from the disc and halo of the MW candidate in realisation $17\_11$. The contribution from the external LG medium (ELGM), denoted as $N_{\rm ELGM}$, includes all material located beyond $R_{200}$ and within a distance of $1000\,$kpc. In contrast, the contribution within the virial radius is referred to as $N_{200}$.

As expected, most of the studied column densities are generally dominated by the disc and halo contribution. In particular, in the case of H\,{\sc i}, Si\,{\sc iii} and O\,{\sc viii}, less than 2\% of the sightlines show a higher ELGM contribution to the column densities compared to that of the disc and halo, with most of them traversing the M31 central region. We note that, while H\,{\sc i} and Si\,{\sc iii} owe this to higher column densities in the disc, O\,{\sc viii} is predominantly present in the hot halo of this MW analogue. Conversely, C\,{\sc iv}, O\,{\sc vi} and O\,{\sc vii} seem to favour ELGM material more than the previously mentioned ions in this simulation, reaching higher column densities beyond $R_{200}$ than in the halo (that is, $N_{\rm ELGM} > N_{200}$) for $16\%, 27\%$ and $9.8\%$ of the sightlines, respectively. In general, we find an excess of extragalactic material in the direction of the barycentre as approximately $9\%$, $15\%$ and up to $40\%$ of all the sightlines with $N_{\rm ELGM}>N_{200}$ tend to cross through the M31 halo in the cases of C\,{\sc iv}, O\,{\sc vi} and O\,{\sc vii}, respectively. This should be considered alongside the fact that the M31 halo covers only about $4\%$ of the sky, indicating that a non-negligible fraction of the ELGM-dominated lines are primarily influenced by the disc and halo gas of the M31. 

In Fig.~\ref{fig:CIV_ratio_mollweide}, we show the ratio between $N_{\rm ELGM}$ and ${N_{200}}$ for C\,{\sc iv} from the point of view of the simulated LSR. The barycentre and anti-barycentre regions are highlighted by the dashed lines. Green zones represent areas of the sky where $N_{\rm ELGM} > N_{200}$, meaning that it would be more likely for an observer to detect ELGM gas in these directions of the simulated sky. The excess of ELGM material is especially evident towards the barycentre direction, which essentially coincides with the position of M31. This indicates that these lines tend to probe material outside the MW's CGM that is mainly related to the M31 gaseous halo and its surroundings. Notably, an excess of ELGM material is also seen in the anti-barycentre direction. This is further supported by Fig.~\ref{fig:LG_ion_distribution}, which shows a tail in the C\,{\sc iv} distribution extending towards the outskirts of the LG. In this direction, the $N_{\rm ELGM}/N_{200}$ ratio does not reach values as high as those associated with certain lines tracing the M31 halo. Qualitatively similar results are obtained for O\,{\sc vi}, in line with the observations of \cite{Sembach03}, which report high O\,{\sc vi} column densities in the general direction of the barycentre. In the case of O\,{\sc vii} and O\,{\sc viii}, the highly ionised CGM of our model MW produces less ELGM-dominated lines in comparison to O\,{\sc vi}, as the CGM contribution to column density for these ions is relatively high. Interestingly, these results are in line with the observational findings of \cite{Bouma19} (but see also \citealt{Richter17}), who determined that a fraction of the material probed by sightlines pointing towards the barycentre direction might correspond to material outside the virial radius of our Galaxy.

Considering the previous results, it is evident that, in order to probe IGrM material, it is necessary to remove from the synthetic lines the high column densities associated with the disc and, to a lesser extent, the CGM of the simulated MW. Fig.~\ref{fig:N_ratio_hist} shows the contributions of the ELGM to the sightlines in realisation $17\_11$, relative to those from the CGM, after applying the filtering method described in Sec.~\ref{sec:filtering} to mask out disc-like motion for the six different ionic species considered in this work (green histograms). Also shown are the non-filtered counterparts including all material inside and outside $R_{200}$ (purple histograms). After the filtering procedure is applied, the non-filtered distributions shift to the right with low ions showing the greatest difference, which can span several orders of magnitude (see upper row of Fig.~\ref{fig:N_ratio_hist}). Conversely, oxygen high ions show a more moderate shift of the distributions. 

These results are expected as low ions tend to better trace the inner galactic regions. In fact, the magnitude of the observed distribution shift gradually decreases from left to right, starting in the upper left panel, demonstrating the impact of the filter on each ion. As a result of the shift, the ELGM contribution to column densities is greatly enhanced, reaching higher column densities beyond $R_{200}$. Specifically, after excluding material consistent with disc motion, we obtain that $N_{\rm ELGM} > N_{200}$ for 81\% of the lines of sight for H\,{\sc i}, 50\% for Si\,{\sc iii}\footnote{In realisation $17\_11$, a significant fraction of Si\,{\sc iii} lines of sight have $N_{200} \approx 0$ after removing disc material. To avoid divergences in the ratio $N_{\rm ELGM}/N_{200}$, we excluded them from our analysis. However, if one takes these lines into account, $f_{\rm ELGM}$ can be as large as $77\%$.}, 76\% for C\,{\sc iv}, 84\% for O\,{\sc vi}, 57\% for O\,{\sc vii}, and 11\% for O\,{\sc viii}. In particular, O\,{\sc viii} and, to a lesser extent, O\,{\sc vii} are found more frequently within the halo, thus resulting in CGM-dominated column densities, even after removing the disc-like contribution. The same global trends are also valid in the case of the $9\_18$ realisation. One might expect these high oxygen ions to be strong tracers of the IGrM. However, in galaxies with hot haloes like the MW, the CGM can dominate, often exceeding the IGrM column density contribution of higher ions across most sightlines.

\begin{figure}   
 \includegraphics[width=.9\columnwidth]{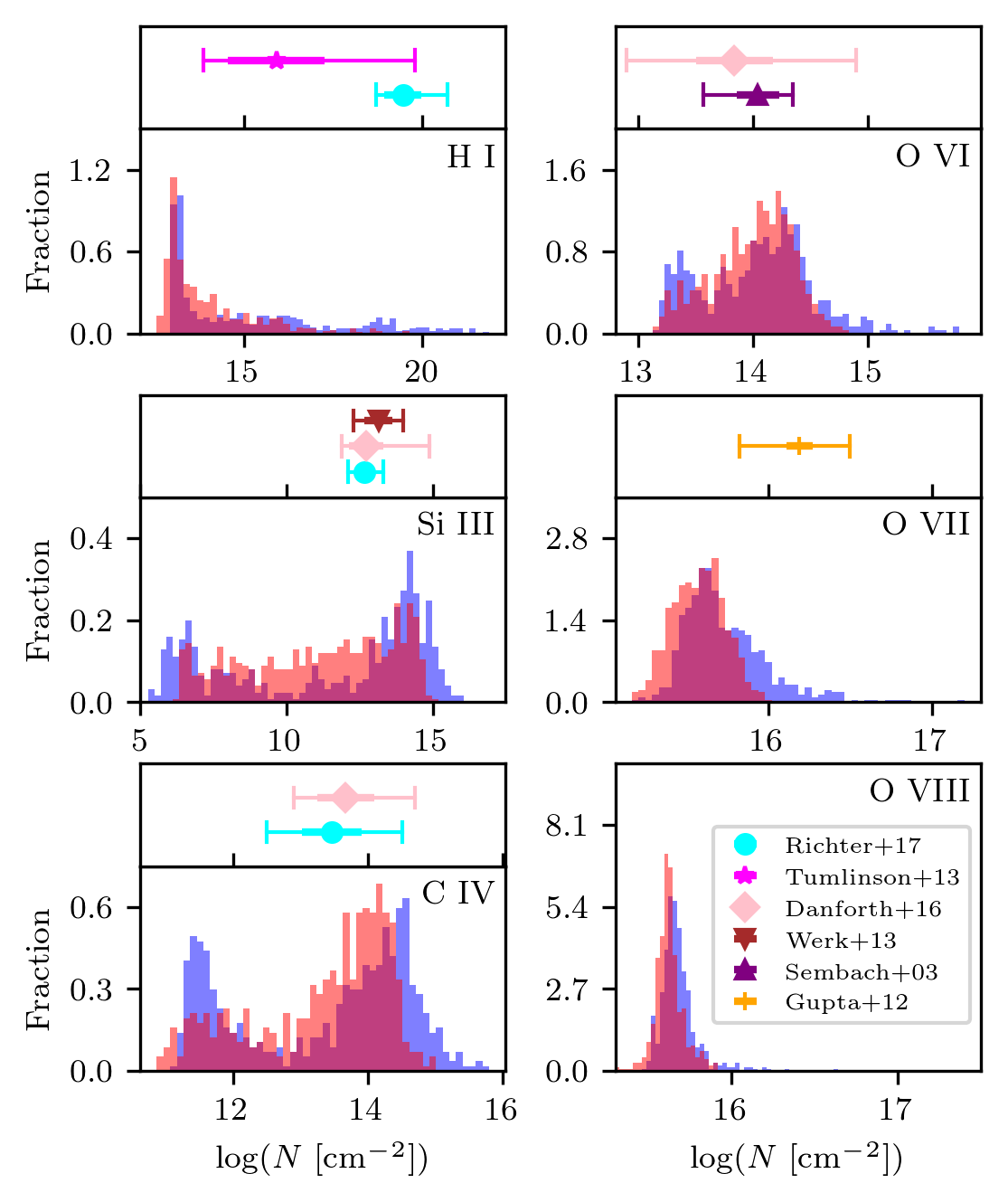}
    \caption{Column density distributions for the six different ions studied in realisation $17\_11$, excluding material with motion consistent with disc rotation. Distributions for sightlines in the barycentre and anti-barycentre direction are shown in blue and red, respectively. The error bars in the upper panels indicate mean and standard deviation, with whiskers extending up to minimum and maximum observed values in the MW halo, external galaxies and the low-redshift intergalactic medium (see text).}
    \label{fig:column_density_hist}
\end{figure}

In Fig. \ref{fig:column_density_hist} we show column density distributions for the six chemical species studied in both barycentre (blue) and anti-barycentre (red) directions, excluding material with motion consistent with disc rotation for realisation $17\_11$. The error bars indicate column density mean and standard deviation, with whiskers extending up to minimum and maximum values in which these ions have been observed in the MW halo \citep[][]{Sembach03, Gupta12, Richter17}, in external galaxies of similar stellar mass \citep{Werk13, Tumlinson13} and in the low-redshift intergalactic medium \citep{Danforth16}. There seems to be no striking differences between the two directions in this regard, although blue histograms generally show higher column density tails. Si\,{\sc iii} and C\,{\sc iv} display two distinct peaks in their distributions, the lower one linked to sightlines with column densities typical of intergalactic gas, which is less contaminated. The second peak, at higher column densities, corresponds to enriched gas in the vicinities of galaxies, with denser environments. In this realisation, the M31 halo contains a relatively high concentration of Si\,{\sc iii} and  C\,{\sc iv} in comparison to the MW (see upper row in Fig.~\ref{fig:LG_ion_distribution}), owing to the strong MW outflows. As a result, some sightlines exhibit significantly lower column densities, as they miss the high-density regions of the M31 halo. In contrast, high oxygen ions do not display this bimodal behaviour, as they primarily trace the hot galactic halo and/or intergalactic gas.

\subsubsection{Covering fraction profiles}

\begin{figure}   
    \includegraphics[width=.9\columnwidth]{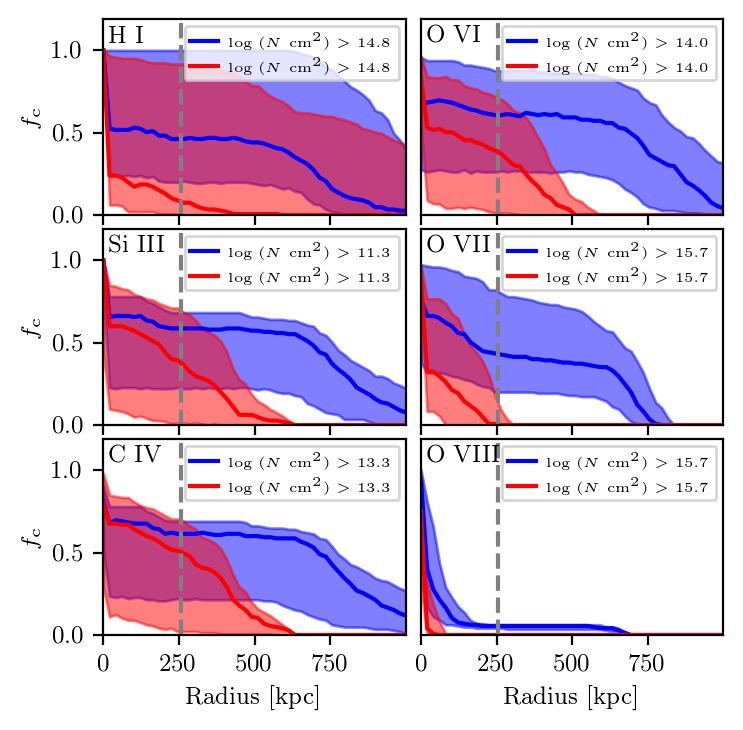}
  \caption{Covering the fraction radial profile for the six studied ions among sightlines evenly distributed in the barycentre (blue) and anti-barycentre (red) directions for realisation $17\_11$. The dashed vertical line marks a distance equal to $R_{200}$ from the observer.}
    \label{fig:fc_profile}
\end{figure}

Fig.~\ref{fig:fc_profile} shows the covering fraction profile from the Sun's location for all ions in realisation $17\_11$ both for the barycentre (red) and anti-barycentre (blue) directions. To build these profiles we placed the observer at the Sun's position and pixelated the simulated sky using the {\sc Healpy} Python package. As already mentioned, sightlines are circumscribed within $20^{\circ}$ of the analysed directions. We then radially binned material along all sightlines and computed column densities at a given radius using only material lying beyond that radius and up to the edge of the high-resolution region at about $1400$~kpc from the centre of the simulated sphere. From these column densities, we finally calculated the covering fraction of each species exceeding the corresponding mean column density, $N_{\rm lim}$, as obtained from their column density distributions. Additionally, we also computed the $\pm1\sigma$ standard deviation error bands, which we indicate as shaded regions in each panel. 

For all studied species, covering fraction profiles decline more slowly when looking to the barycentre than to the anti-barycentre direction. Towards the later, the covering fractions decrease faster, reaching negligible values well outside the MW halo at around $500\,$kpc for all species with the exception of the highest oxygen ions. This is expected as there should be more material towards the LG barycentre owing to the presence of M31 and its CGM \citep[see, e.g.][]{Nuza14a,Biaus22}. In particular, almost all species, with the exception of O\,{\sc viii} and, to a certain extent, O\,{\sc vii}, roughly cover half of the beam ($f_{\rm c}\sim0.5$) in the barycentre direction for column densities higher than the mean value, independent of the distance to the Sun. This indicates that the mean column density values in both the MW and M31 haloes are similar for these species, although, as clearly seen in Fig.~\ref{fig:column_density_hist}, the corresponding barycentre distributions display larger column density tails in comparison to the anti-barycentre ones.

Covering fractions for the highest oxygen ions decline faster than low ions, suggesting that, in this particular simulation, the MW halo mainly determines the column density limit applied. This is true because, as already mentioned, most of the O\,{\sc viii} in the simulation originates in the large-scale outflows seen in the MW candidate at $z=0$. Finally, we note that taking the mean or the median column density along each sightline does not significantly change $N_{\rm lim}$, other than in the case of the more clustered ions, but without qualitatively changing the results.

\begin{figure*}
    \centering
    \includegraphics[width=1.6\columnwidth]{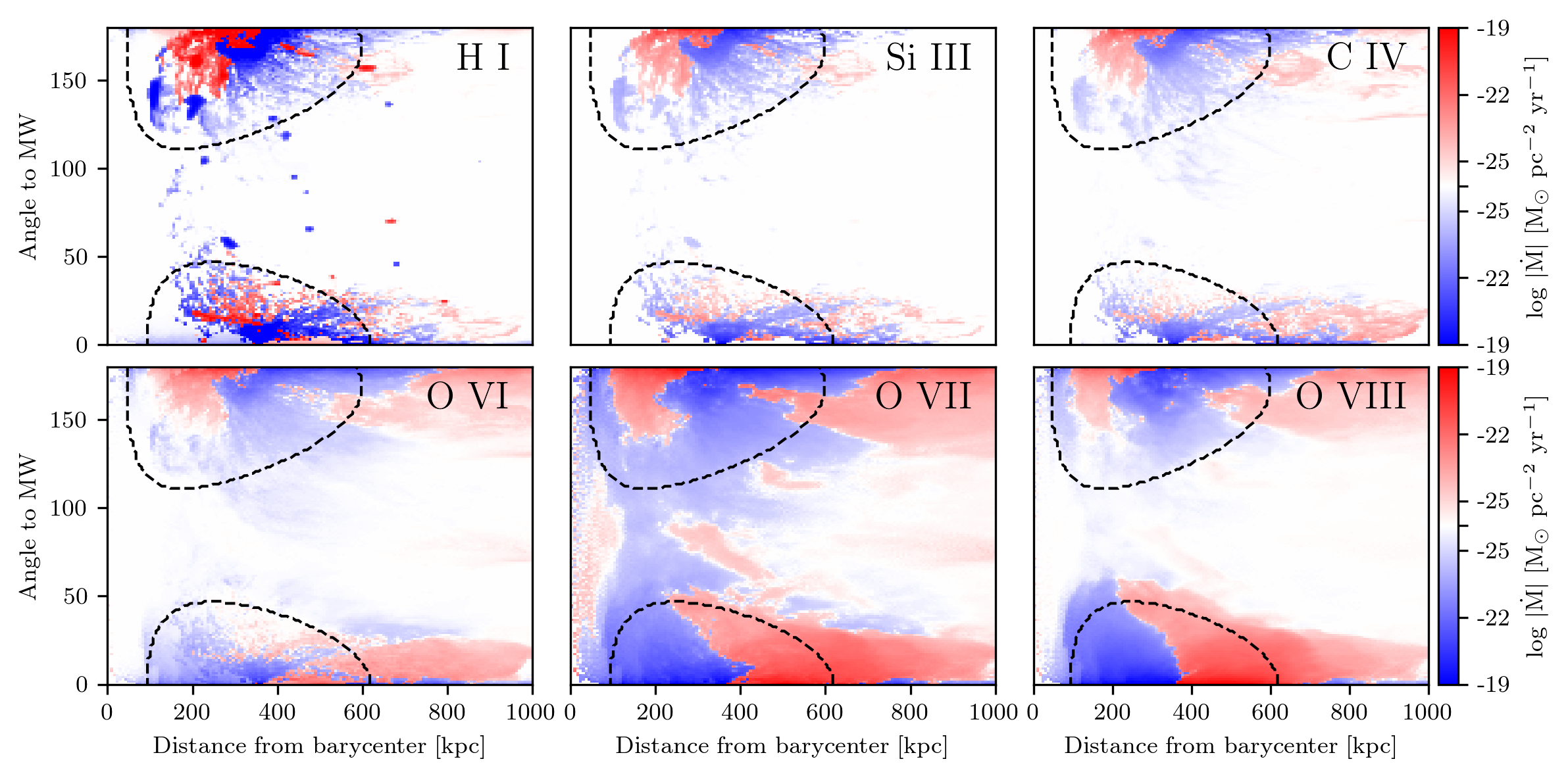}
    \caption{Radial mass flux as seen from the Local Group barycentre for each ion. The position of each point in space with respect to the barycentre is measured relative to the MW. Negative flux (blue) represents gas moving towards the barycentre and positive flux (red) represents material moving away from it. Note that the numbers in the colourbars represent the symmetric exponent of the mass flux in the inflow and outflow cases. In all panels, black dashed lines enclose the circumgalactic medium of the MW and M31 up to the corresponding $R_{200}$ radii.
    }
    \label{fig:LG_flow}
\end{figure*}

\subsection{Chemically enriched gas kinematics}
\label{sec:chem_kin}

To gain insight on the global motion of enriched material within simulations, we analyse the mass flux of each ion relative to the barycentre of the simulated LG. For each realisation, we compute the barycentre’s position and velocity including all matter (gas, stars and DM) located within a radius of $1000\,$kpc from the midpoint between the MW and M31 candidates. As it was shown in \cite{Biaus22}, this distance roughly defines the edge of the LG as it corresponds to the transition between the local and Hubble velocity flows. At larger distances, the Hubble flow takes over, and the GSR radial velocity of gas and galaxies gradually aligns with the universal expansion.

Fig. \ref{fig:LG_flow} shows the distance and angular distribution of radial mass inflows of each species with respect to the LG barycentre in the $17\_11$ realisation, where angles are measured relative to the MW's position. Negative fluxes (colour-coded in blue) indicate infalling gas towards the barycentre, whereas positive fluxes (red) stand for receding material. Dashed lines enclose the CGM of the MW (centred at $0^{\circ}$) and M31 (centred near $180^{\circ}$) up to $R_{200}$. The overall gas kinematics shows a global motion towards the LG's barycentre with, however, some material flowing outwards from the main haloes as a result of strong outflows in the MW and/or accretion patterns within the virial radius of M31. This is owing to the fact that, in realisation $17\_11$, the two main galaxies show distinctive features at $z=0$, with the MW experiencing strong outflows and M31 accreting material as a result of its large mass. In particular, most of the H\,{\sc i} kinematics is concentrated in regions associated to galaxies, as expected for an ion tracing cold gas. Within the two main haloes signs of H\,{\sc i} accretion can be seen, and small galaxies can be traced by their H\,{\sc i} mass flux outside the MW and M31, most of them falling towards the barycentre. Si\,{\sc iii} kinematics are dominated by the two main galaxies, extending up to $R_{200}$, but failing to trace the kinematics of smaller galaxies as good as H\,{\sc i}. This is reasonable as smaller galaxies are expected to be less chemically enriched. C\,{\sc iv} shows similar kinematic patterns than that of Si\,{\sc iii}, though its mass flux extends farther into the outer parts of the LG owing to its higher ionisation potential tracing hotter material in the external regions of galactic haloes.

Among the hot-tracing oxygen high ions, the MW shows very strong outflows, particularly in O\,{\sc vii} and O\,{\sc viii}, which results in a significant mass flux beyond $R_{200}$ showing that galactic haloes are responsible for polluting the intergalactic medium with heavier elements. On the contrary, hot gas in the M31 halo shows movement towards the galaxy's centre. In both directions and beyond the two main galaxies, a tail of hot gas outflowing from the LG is evident in all three oxygen ions. Interestingly, O\,{\sc vii} is the only one among the studied ions that shows a significant mass flux in the direction perpendicular to that of the two main galaxies and, although most of this flux has negative velocities (i.e. this material is falling towards the LG barycentre), there is also some outflowing material. This is also seen in realisation $9\_18$ for this particular ion.

\subsection{Modelling observed ion kinematics}

In this section, we study the gas kinematics of different ions in the LG realisation $17\_11$ from the usually adopted reference frames to mimic observed distributions. As explained above, the observer is located in the midplane of the MW candidate at a galactocentric distance of $r=8\,$kpc assuming that the velocity of the Sun is purely tangential, which defines the model LSR. From this simulated reference frame, it is possible to further construct the GSR and LGSR analogues (see Section~\ref{sec:reference} for details).

\begin{figure*}
    \centering
    \includegraphics[width=1.5\columnwidth]{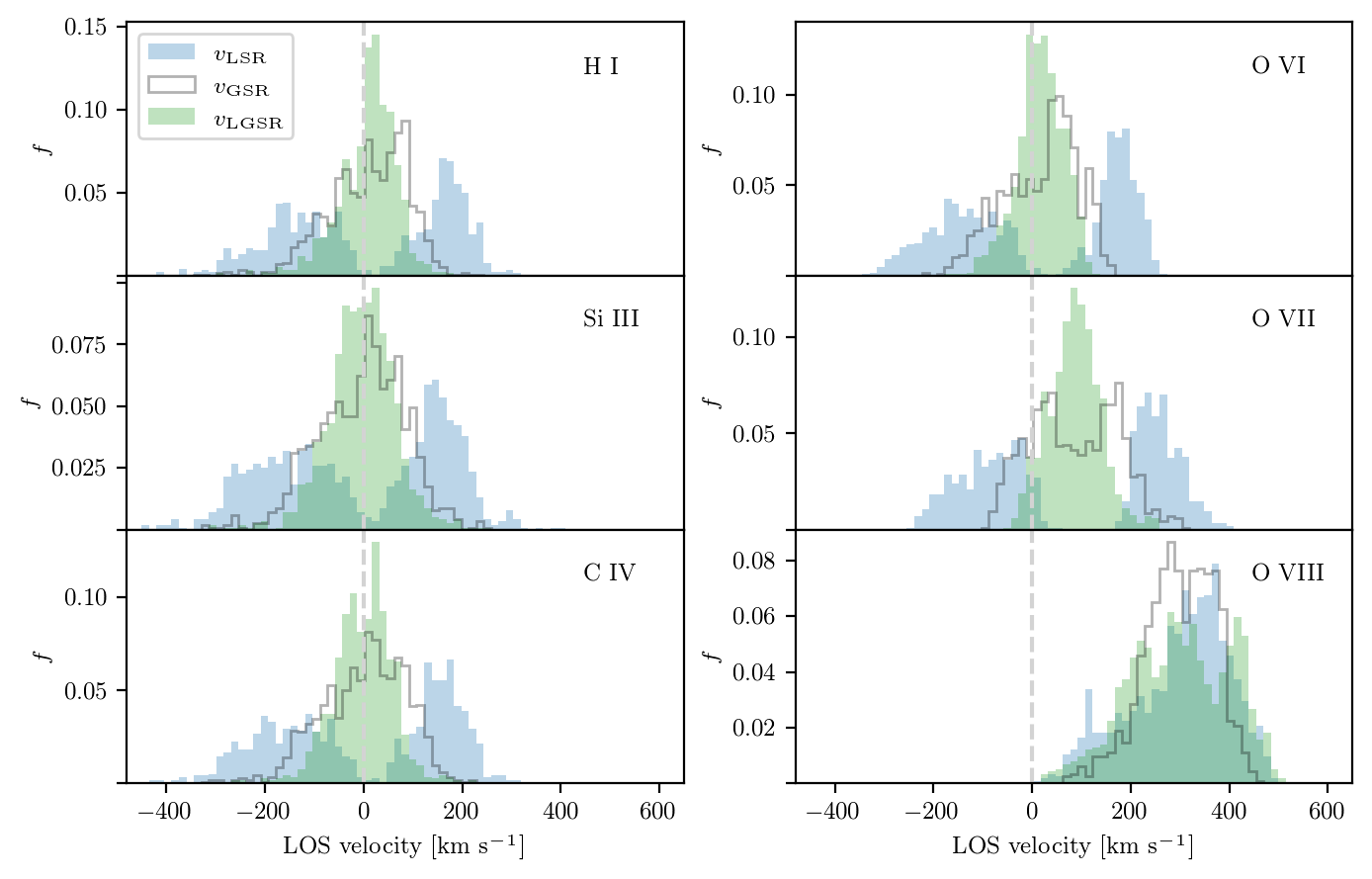}
    \caption{Mass-weighted velocity distributions for a combined set of 1154 lines of sight (LOS) pointing to the barycentre and anti-barycentre directions for each ionic species in realisation $17\_11$. Mock observed distributions from different reference frames are shown: $v_{\rm LSR}$ (blue), $v_{\rm GSR}$ (empty) and $v_{\rm LGSR}$ (green).}
    
    \label{fig:v_histograms}
\end{figure*}

In Fig. \ref{fig:v_histograms}, we show the velocity distribution for 577 sightlines evenly distributed within the barycentre and anti-barycentre cones, excluding all material consistent with disc rotation. In this way, we are able to construct an homogeneous distribution of sightlines that combine the information of the two preferred directions simultaneously. Histograms for each of our six studied species are represented as they appear in the simulated LSR, GSR and LGSR reference frames. It is evident that all distributions narrow down when transforming from the LSR to the GSR, narrowing even further when transforming from the GSR to the LGSR, the only exception being the distribution of O\,{\sc viii} which will be discussed below. This is consistent with the expected widening of the distributions when moving from the LGSR to the GSR as a result of the MW's peculiar motion with respect to M31. This behaviour is a necessary, although not sufficient, condition if material is falling towards the LG barycentre \citep[e.g.][]{Sembach03,Richter17,Biaus22}. In fact, in \cite{Biaus22}, using both the general barycentre and anti-barycentre directions defined as two sky regions separated by the lines $l = 180^{\circ}$ and $b = 0^{\circ}$, we have shown that the mapped material outside the MW's halo is, in fact, moving towards the LG barycentre in realisation $17\_11$. Moreover, this general trend is consistent with the observed velocity distributions presented in \cite{Richter17} for sightlines towards M31 and its antipode in the sky (see their Fig. 11). In this work, however, we primarily focus on the narrower barycentre/anti-barycentre directions defined by the cones, while still arriving at similar conclusions for most of the analysed ions. When transforming the distributions to the LSR, histograms get wider once again displaying a clear velocity dipole as a result of the rotational velocity of the Sun around the MW.

In general, there seems to be very little differences between the velocity distributions for the three species tracing colder gas (namely H\,{\sc i}, Si\,{\sc iii} and C\,{\sc iv}) as they follow a similar, strongly clustered, gaseous phase within the LG volume. However, when looking at the oxygen high ions, distributions shift towards higher velocity values in all reference frames as the ionisation number grows. This shift is more noticeably in the case of O\,{\sc vii} and O\,{\sc viii} suggesting that, in LG realisation $17\_11$, oxygen high ions mainly trace material flowing outwards from the MW in the form of hot outflows, as already explained. This shows that some of this hot material is outflowing from the LG barycentre as a result of supernova feedback instead of falling in (see the lower panels of Fig.~\ref{fig:LG_flow}). Conversely, the MW candidate in simulation $9\_18$ does not show strong outflows at $z=0$ (see, e.g. Fig. 8 of \citealt{Biaus22} and Fig. \ref{fig:app_v_histograms:918} of Appendix \ref{app:9_18}). Therefore, in that case, oxygen high ions follow similar trends as low ions, displaying the expected decrease of velocity dispersion between different frames for material falling onto the LG barycentre.

\subsection{Ion pressure profiles and kinematics}
\label{sec:chem_pressure}

In \cite{Bouma19}, the authors presented a sample of high-velocity absorbers in the neighbourhood of the general barycentre and anti-barycentre directions. These observations were used to infer gas properties in the location of the absorbers. By comparing the derived gas pressures with the spherically symmetric halo model of \cite{Miller&Bregman15}, they suggest that some of the low-pressure absorbers in the barycentre direction may be located in the outer halo of the MW and, perhaps, even beyond its virial radius. In the anti-barycentre direction, however, the inferred gas pressures suggest that the absorbers are located at smaller distances of about $100\,$kpc from the Sun.

\begin{figure*}
    \centering
    \includegraphics[width=1.4\columnwidth]{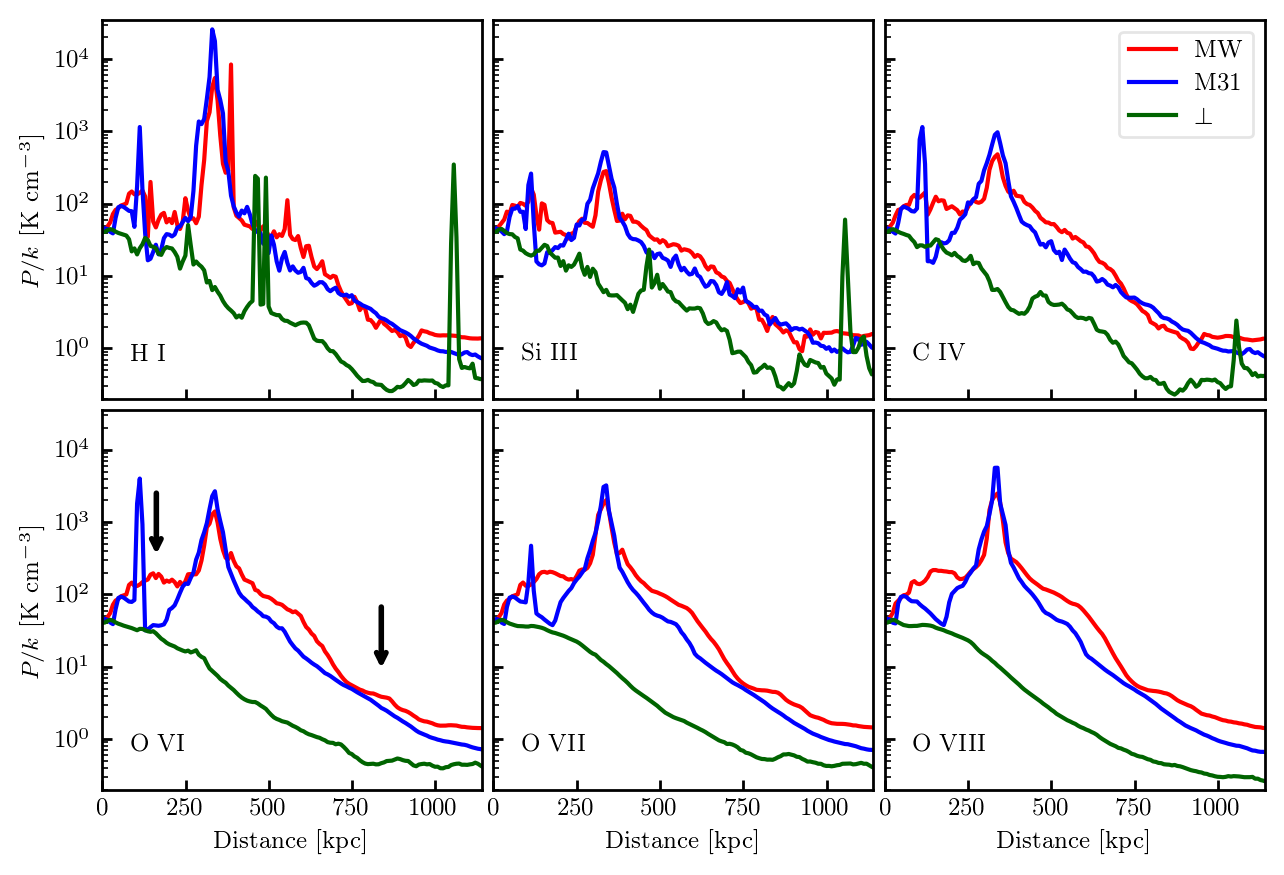}
    \caption{Gas pressure profile of the different ionic species for Local Group realisation $17\_11$ measured from the midpoint between the MW and M31 in different directions. The solid lines show profiles pointing towards the MW (red), M31 (blue) and averaging over many directions perpendicular to the radial vector joining the two galaxies (green). Note that the two peaks seen at $r\sim 300\,$kpc in the red and blue lines correspond to the MW and M31 haloes. Arrows indicate locations in front (left) and behind (right)  the MW candidate.}
    \label{fig:LG_pressure_profile}
\end{figure*}

In this section, we want to assess the distance-pressure relation as obtained with our more realistic LG simulations in order to compare with observed trends. This is important because real galactic halos are typically not spherically symmetric and often exhibit evidence of interactions with other galaxies. In Fig. \ref{fig:LG_pressure_profile}, we show the pressure profile of the LG as measured from the midpoint between the MW and M31 in different directions. The red and blue lines show pressure profiles averaged over cones with $20^{\circ}$ opening angles in the direction of the MW and M31, respectively. The green lines show pressure profiles averaged within $\pm 5^{\circ}$ from the directions perpendicular to the radial vector joining both galaxies. 

Overall, we found that pressure profiles along MW and M31 directions display similar trends. The neutral, strongly clustered H\,{\sc i} component reaches high $P/k$ values of the order of $10^4$~K cm$^{-3}$ within the MW and M31 haloes, but declines to lower values compared to other species beyond the influence of the two main galaxies. This is to be expected as this component is colder than the rest of the ions within the filamentary distribution of material seen in the LG. The presence of MW, M31 and satellite galaxies is revealed by peaks in the pressure profiles. Remarkably, as we shift towards less-clustered, hotter ions, the profiles become smoother, more accurately reflecting the ambient pressure of the IGrM (this is in line with Fig.~3 of \cite{Damle22}, which shows more power on small angular scales for cold ions, essentially tracing the clumpiness of ions). For distances smaller than the location of the MW and M31 haloes ($r\lesssim300\,$kpc), pressure profiles show some dependence with direction, although with a roughly even distribution of IGrM and halo gas. However, at larger distances, pressure anisotropy in the LG becomes evident, with pressure values consistently higher in the direction of MW and M31 compared to perpendicular sightlines, which reflects the absence of MW-sized galactic haloes in these directions. The pressure decline observed perpendicular to the MW-M31 axis may indicate the presence of a `hot bridge' between these galaxies as suggested by \cite{Qu2021}, in line with previous LG simulations \citep{Nuza14a}. This feature is also seen in realisation $09\_18$ (see Fig. \ref{fig:app_LG_pressure_profile_918} in the Appendix \ref{app:9_18}).

In Fig. \ref{fig:pressure_crosses}, we present the mass-weighted, average pressure and velocity values for sightlines in the barycentre and anti-barycentre directions binned in radial bins of $50\,$kpc from the Sun's location in realisation $17\_11$. The left-hand panel shows the simulated LSR velocity values together with the observational estimates of \cite{Bouma19} for their absorber sample. Similarly, the right-hand panel shows the simulated GSR velocity values. As expected, gas pressure decreases with distance for all species in agreement with observations, yet the values themselves do not accurately match the observed ones: pressure values derived in \cite{Bouma19} are generally lower than the simulated ones at the corresponding distance. This disagreement could be caused by multiple factors. First and foremost, the simulated MW has a greater mass ($M_{200}=1.89 \times 10^{12}$~M$_\odot$; see Table~\ref{table:1}) than most observational estimates, which place the mass at about $10^{12}$~M$_\odot$. Recent estimates of the MW mass at $r\lesssim R_{200}$ include that of \cite{Callingham19}, \cite{Shen_2022} and \cite{Makarov25} that report values of $1.17_{-0.15}^{+0.21} \times 10^{12}$~M$_\odot$, $(1.08 \pm 0.11) \times 10^{12}$~M$_\odot$ and $(7.9 \pm 2.3) \times 10^{11}$~M$_\odot$, respectively. This results in higher gas pressures for the simulated MW halo. Moreover, the CGM of the MW in realisation $17\_11$ exhibits strong outflows, indicating intense thermal activity that likely deviates from the actual CGM and could further elevate pressure values. In contrast, in realisation $9\_18$, where the MW candidate has a very similar $M_{200}$ but does not present these strong outflows, pressures are systematically lower (see Fig.~\ref{fig:app_pressure_crosses_9_18} of Appendix~\ref{app:9_18}). The resulting values are, however, of the same order of magnitude than in realisation $17\_11$ and do not closely resemble the observed pressure-distance distribution. This indicates that, in order to fairly reproduce the detailed properties of the observed MW halo, LG simulations should also gauge the variance of the CGM properties given by the different possible instantaneous thermodynamical states of the system.

Note that, in realisation $17\_11$, the very strong outflows in the MW halo push the gas away with $v_{\rm GSR} > 0$ in both directions up to $r\approx 300\,$kpc. This is not the case in realisation $9\_18$, in which most ions show signs of accretion at the innermost parts of the CGM, except for the oxygen highest ions, which trace hotter gas in the galactic halo. For large distances ($d \gtrsim 300$~kpc), gas velocities in the GSR are negative in the barycentre direction and positive in the anti-barycentre direction.  This is consistent with the idea that the MW seems to be ramming into LG gas in the barycentre direction ($v_{\rm GSR} < 0$), while gas lying towards the anti-barycentre direction (the outskirts of the LG) tends to lag behind ($v_{\rm GSR} > 0$). This effect is much less pronounced in the case of realisation $9\_18$, primarily owing to the lower relative velocity between MW and M31. Furthermore, gas peculiar motions in this case tend to wash out the signal in the GSR frame.

\begin{figure*}
    \centering
    \includegraphics[width=.8\columnwidth]{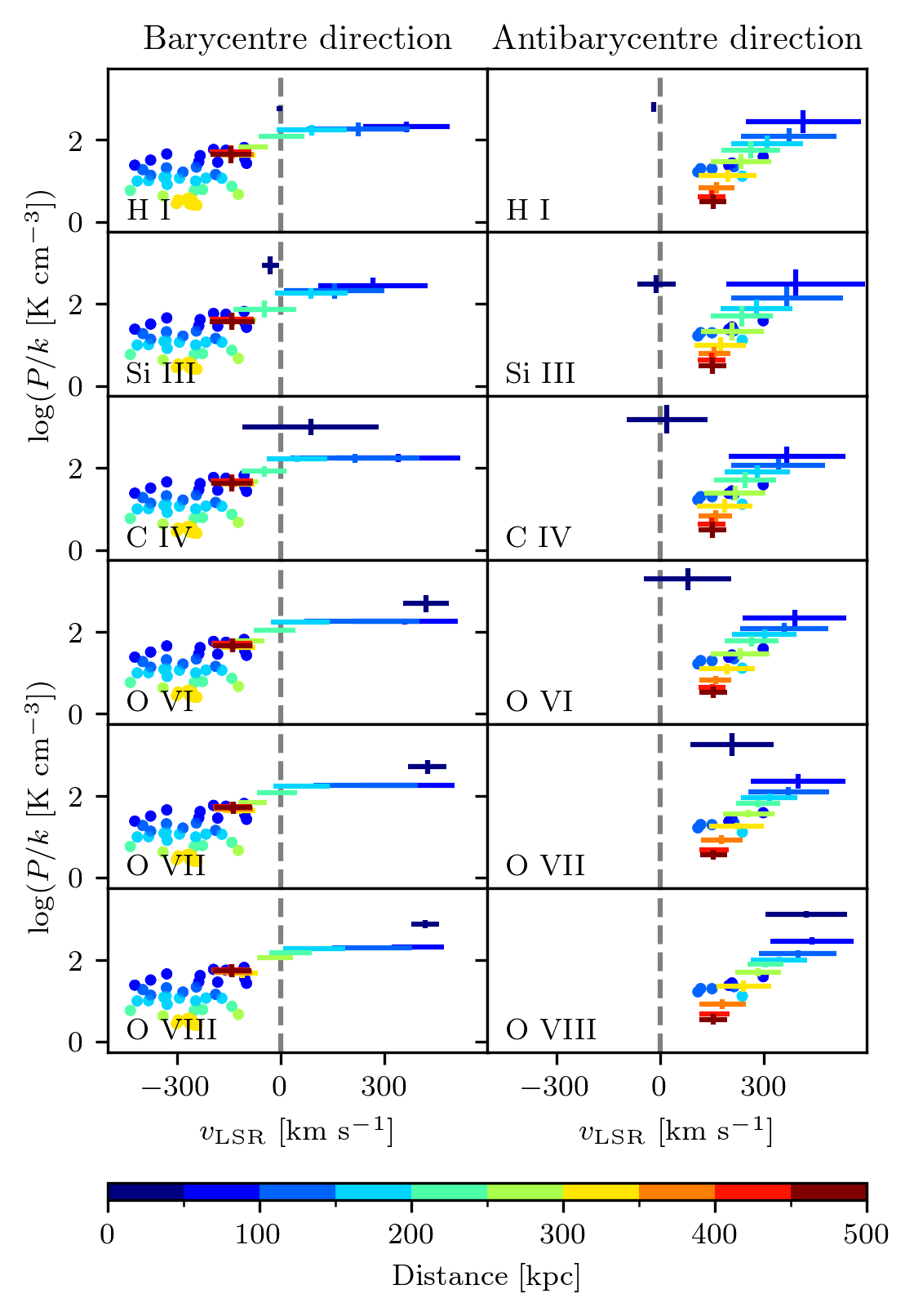}
    \includegraphics[width=.8\columnwidth]{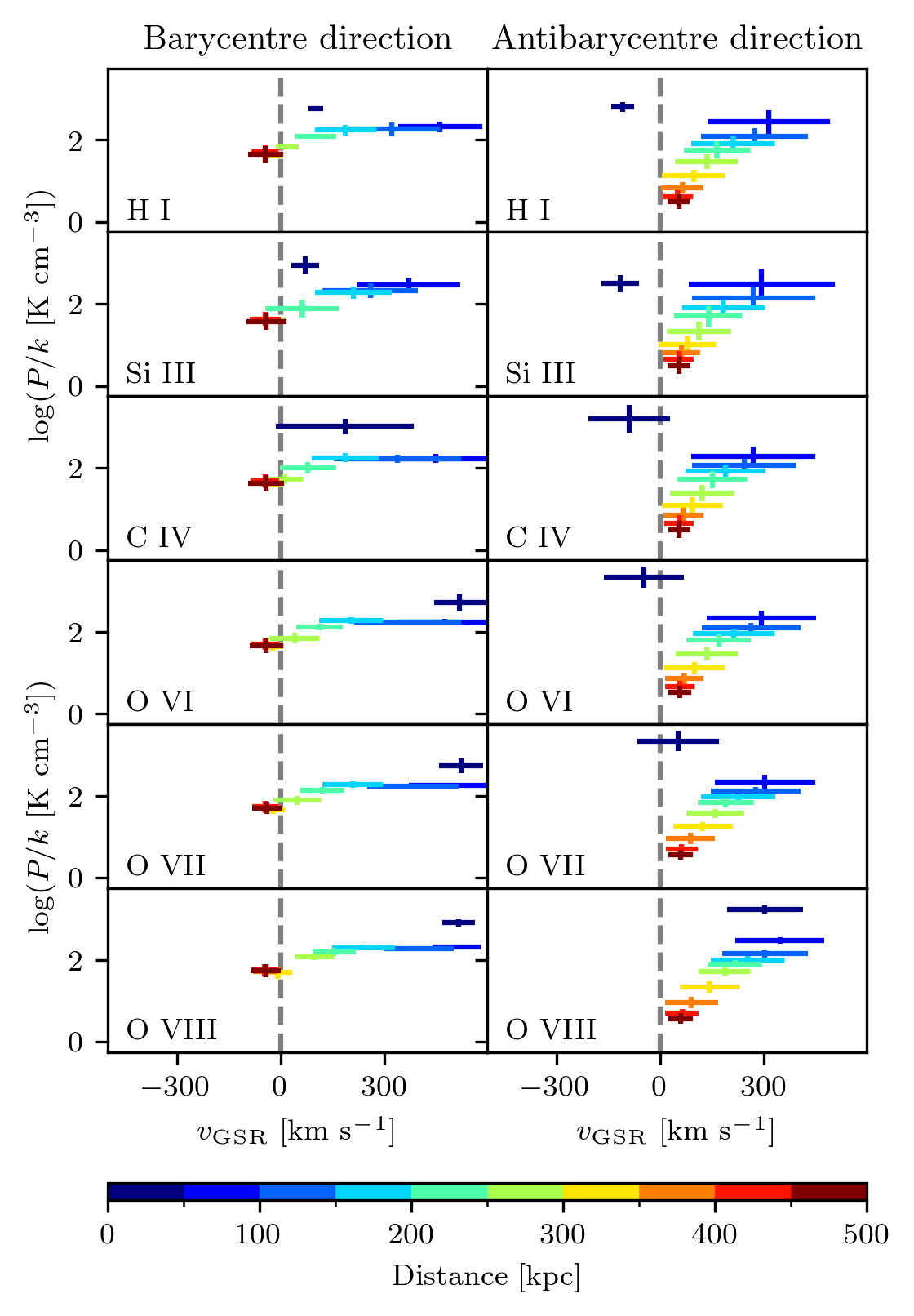}
    \caption{Distance-binned pressure and line-of-sight velocity averages for realisation $17\_11$ in the barycentre and anti-barycentre directions for the LSR (left panel) and the GSR (right panel) reference frames. Overplotted as dots in the left panel are the $v_{\rm LSR}$ and $P/k$ values derived in \cite{Bouma19} for the absorber selection studied in that work.
    }
    \label{fig:pressure_crosses}
\end{figure*}

Regarding the dependency of these trends with direction, the figures show that gas in the anti-barycentre direction in both simulations exhibits lower pressure values than in the barycentre at distances beyond $R_{200}$. Higher pressure values towards the barycentre are expected, as gas in the opposite direction is located towards the outskirts of the LG, where lower densities dominate. Interestingly, the same trend can also be seen in Fig.~\ref{fig:LG_pressure_profile} from the point of view of the MW (red solid line), especially for high oxygen ions. In this plot, the MW is located at a distance $d\sim 300\,$kpc from the LG midpoint and the barycentre and anti-barycentre directions point towards the left and the right of the MW pressure peak, respectively. At large $d$, the red solid line shows lower pressures than the corresponding blue solid line at $d\lesssim 300\,$kpc (see arrows). This suggests a potential observational bias in the absorber sample of the HST/COS Legacy Survey of HVCs used in \cite{Bouma19}, where lower-pressure absorbers are mainly observed towards the barycentre. In contrast, our simulations consistently show higher pressures in the barycentre direction for gas outside the MW halo, implying that observed lower-pressure absorbers may be overrepresented in this general direction. This raises potential concerns about directly applying isotropic pressure models, such as the one described in \cite{Miller&Bregman15}, to interpret observational data.

\section{Discussion}
\label{sec:discussion}

To mask out gas belonging to the MW disc in the simulated sightlines, we applied the filtering method of \cite{Westmeier18}. This approach is employed to mimic one of the usual observational techniques designed to separate the contribution of the galactic disc from that of the CGM in HVC absorbers.As shown in Section~\ref{sec:filtering}, the filtering method effectively removes disc gas, leaving only material inconsistent with galactic rotation. This can be seen in Fig. \ref{fig:N_ratio_hist}, where we applied this separation to our gas cells, resulting in a much greater proportion of sightlines being ELGM-dominated in relation to their typical column densities. This supports the effectiveness of our filtering method for probing gas beyond the MW halo. However, it is also possible that some of the excluded material may also originate from the CGM, although this could happen both in observed and simulated samples. Moreover, owing to the nature of the method, it can also introduce a bias in the velocity range of the non-disc material being analysed. In this case, gas with disc-like velocities could be masked out, even if it belongs to the CGM or IGrM.

One additional aspect that needs to be considered further when comparing MW or LG cosmological simulations to absorber data is the thermodynamic state of the simulated CGM at $z=0$. The latter will naturally exert a strong influence on the ionisation state of the different species, especially in the CGM. This is demonstrated from the variation observed between the two MW candidates in realisations $9\_18$ and $17\_11$. In the former, gas accretion is significantly enhanced, whereas in the latter, strong supernova-driven outflows heat and ionise the material within the galactic halo, producing strong outflows that pollute the IGrM \citep[][]{Biaus22}. In this work, we have examined these differences by analysing the kinematic properties of the different chemical species within the MW halo in both realisations (see Sections~\ref{sec:chem_kin} and~\ref{sec:chem_pressure}). Future LG simulations should also take these factors into account.

When considering the overall kinematics, we have shown that our simulations support a scenario in which the MW moves through LG gas, dragging its CGM along, while approaching gaseous material in the barycentre direction and receding from IGrM gas in the anti-barycentre one. Regarding the direction-dependent anisotropy seen in the pressure-velocity plane, we find that pressure values for the different species are consistently lower in the anti-barycentre direction at distances beyond $R_{200}$. Therefore, the absence of low-pressure absorbers (i.e. located outside the MW halo) in the HST/COS Legacy Survey of HVCs towards the general anti-barycentre region, as reported by \cite{Bouma19}, might indicate a potential observational bias in their sample, preferentially selecting higher-pressure systems inside the CGM behind the direction of the MW's global motion within the LG.

\section{Summary}

We have presented a follow-up study of \cite{Biaus22}, our previous paper in this series, that focuses on the analysis of kinematic properties of gas and galaxies using LG simulations that belong to  the {\sc Hestia} project. The latter consists of a suite of LG-like regions reproducing MW and M31 analogues and their satellites at small scales, as well as their main cosmographic features at large scales. We extended our analysis to include the study of  six different ionic species in the gaseous distribution (H\,{\sc i}, Si\,{\sc iii}, C\,{\sc iv}, O\,{\sc vi}, O\,{\sc vii,} and O\,{\sc viii}), mainly focusing on their distribution (e.g. column densities and covering fractions), kinematics (LSR, GSR, and LGSR velocity distributions), and pressure profiles in the high-resolution LG {\sc Hestia} realisation $17\_11$. When needed, we complemented our analysis with the high-resolution realisation $9\_18$ from the same suite. 

To separate the contribution of rotating gas that belongs to the galactic disc of the MW candidate from the CGM and IGrM counterparts, we implemented the filtering procedure presented in \cite{Westmeier18}, which describes a traditional method to separate the disc rotating gas from the total signal using simple geometric and kinematic considerations. Overall, the two simulations show the same global trends for material outside the galactic haloes, but they present some differences on the CGM thermodynamical properties of each MW candidate. 

The main results can be summarised as follows:

\begin{itemize}
    \item We find that the CGM of the MW leaves strong imprints on the observed kinematic patterns. CGM column densities are high in the case of cold gas-tracing ions, while galactic outflows can have much higher velocities than the ones predicted for IGrM gas. Galactic outflows also heat up the CGM, making higher oxygen ions more abundant in the halo region, reducing their potential as IGrM-tracking species. 
    
   \item An excess of ELGM-dominated lines (e.g. C\,{\sc iv} and O\,{\sc vi}) in the general direction of the barycentre is observed, owing partly to the presence of the M31 halo, but also to an overall denser IGrM in this direction. This is in line with the results of \cite{Bouma19}, who reported that potential LG absorbers were present only towards the barycentre.

    \item Cold gas-tracing ions (H\,{\sc i}, Si\,{\sc iii,} and C\,{\sc iv}) have kinematics dominated by disc and CGM gas. Most of their mass flux with respect to the LG barycentre is concentrated within the CGM of galaxies. Using geometric and kinematic assumptions to filter out disc material, however, does enhance the detectability of the IGrM. The exclusion of gas with velocities consistent with galactic rotation results in at least more than $50\%$ of the sightlines dominated by ELGM column densities for H\,{\sc i}, Si\,{\sc iii}, C\,{\sc iv,} and O\,{\sc vi}.
    
    \item Hot gas-tracing ions (O\,{\sc vi}, O\,{\sc vii,} and O\,{\sc viii}) have kinematics that extend into the CGM and the IGrM. In our simulations,  O\,{\sc vii} and O\,{\sc viii} predominantly trace CGM outflows, with O\,{\sc vii} exhibiting substantial mass flux beyond the CGMs of MW and M31. The high degree of ionisation in the MW halo (partly due to its higher temperature compared to the IGrM) results in a smaller fraction of sightlines with ELGM-dominated column densities for O\,{\sc vii} and O\,{\sc viii} in comparison to the other ions studied in this work.

    \item Gas pressure profiles, from the MW's point of view, show systematically higher values across all studied ions towards the LG barycentre in contrast to its antipode in the sky. This also supports the interpretation that IGrM absorbers are more likely to be detected along this general direction.

\end{itemize}
    
Our results support a general scenario in which the MW rams into material near the LG barycentre, while receding from gas in the opposite direction. This picture aligns with the observed $v_{\rm LSR}$ dipole in the barycentre--anti-barycentre direction. However, our simulations show that the disc-filtering method and the peculiarities of each LG realisation could introduce biases that might affect the interpretation of observations. In this sense, a larger absorber catalogue is needed to properly map the IGrM.
    
\section*{Data Availability}

The scripts and plots for this article will be shared on reasonable request to the corresponding author. The {\sc Arepo} code is publicly available \citep{Weinberger20}.    
    
\begin{acknowledgements}
SEN and CS are members of the Carrera del Investigador Cient\'{\i}fico of CONICET. They acknowledge funding from Agencia Nacional de Promoci\'on Cient\'{\i}fica y Tecnol\'ogica (PICT-2021-0667).

\end{acknowledgements}

%%%%%%%%%%%%%%%%%%%%%%%%%%%%%%%%%%%%%%%%%%%%%%%%%%

\bibliographystyle{aa}
\bibliography{biblio}

\newpage

\begin{appendix}
\onecolumn

\section{Kinematics and pressure values in realisation $9\_18$}
\label{app:9_18}

\begin{figure*}[h!]
    \centering
    \includegraphics[width=.75\columnwidth]{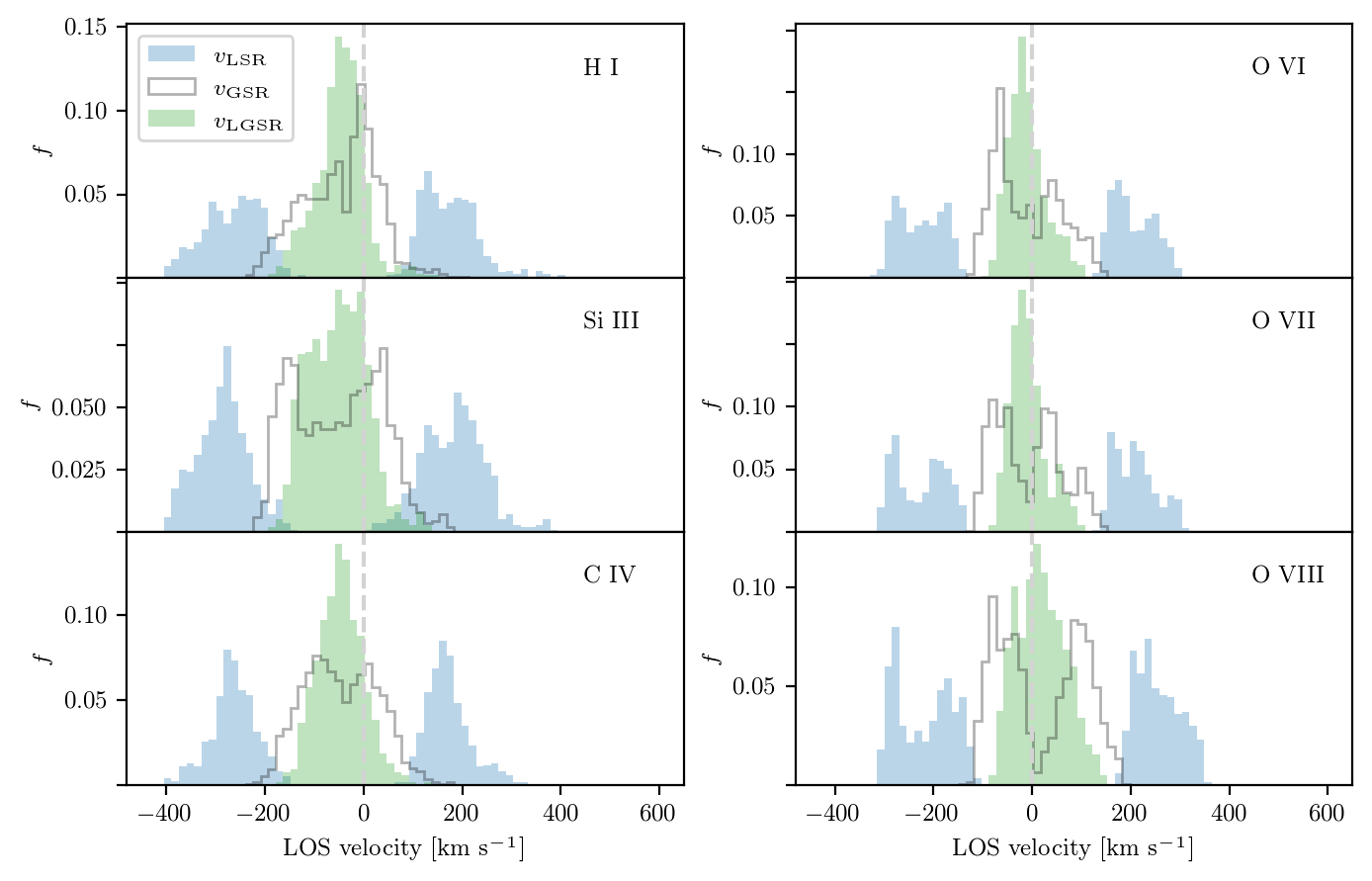}
    \caption{Mass-weighted velocity distributions for a combined set of 1154 lines of sight (LOS) pointing to the barycentre and anti-barycentre directions for each ionic species in realisation $9\_18$. Mock observed distributions from different reference frames are shown: $v_{\rm LSR}$ (blue), $v_{\rm GSR}$ (empty) and $v_{\rm LGSR}$ (green).}
    \label{fig:app_v_histograms:918}
\end{figure*}

In Fig. \ref{fig:app_v_histograms:918}, we show the gas velocity distribution of different species in the LSR, GSR and LGSR frames for 577 sightlines evenly distributed within the barycentre and anti-barycentre cones in realisation $9\_18$, excluding all material consistent with disc rotation. In contrast with realisation $17\_11$ (see Fig. \ref{fig:v_histograms}), distributions for the high oxygen ions are centred at $v \approx 0$, as there are no prominent hot outflows dominating the kinematics of the CGM at $z=0$ in this simulation.

\begin{figure*}[h!]
    \centering
    \includegraphics[width=.75\columnwidth]{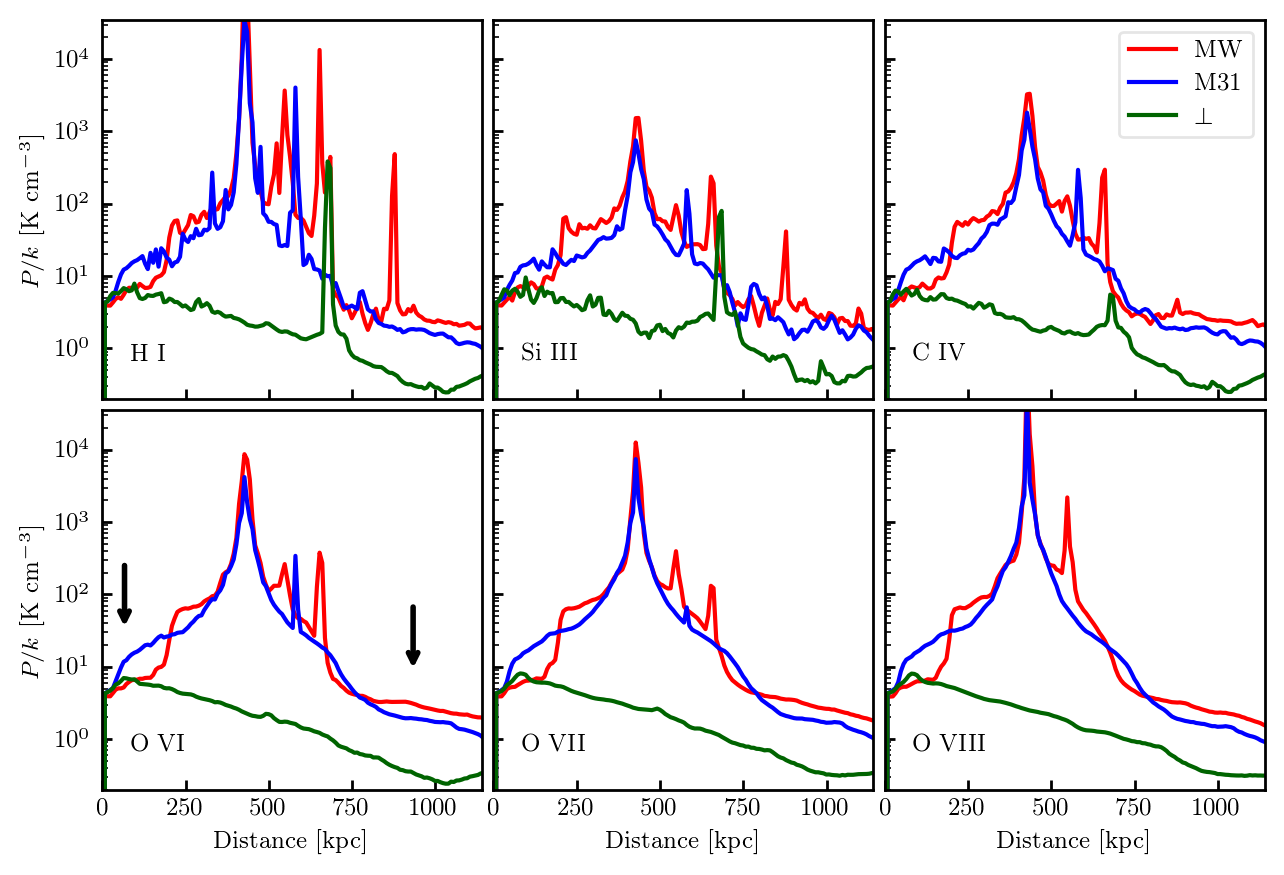}
    \caption{Gas pressure profile of the different ionic species for Local Group realisation $9\_18$ measured from the midpoint between the MW and M31 in different directions. The solid lines show profiles pointing towards the MW (red), M31 (blue) and averaging over many directions perpendicular to the radial vector joining the two galaxies (green). Note that the two peaks seen at $r\sim 400\,$kpc in the red and blue lines correspond to the MW and M31 haloes. Arrows indicate locations in front (left) and behind (right)  the MW candidate.}
    \label{fig:app_LG_pressure_profile_918}
\end{figure*}

In Fig. \ref{fig:app_LG_pressure_profile_918}, we show the pressure profile of the LG as measured from the midpoint between the MW and M31 in different directions for realisation $9\_18$. The red and blue lines show pressure profiles averaged over cones with $20^{\circ}$ opening angles in the direction of the MW and M31, respectively. The green lines show pressure profiles averaged within $\pm 10^{\circ}$ from directions perpendicular to the radial vector joining both galaxies. In this regard, realisation $9\_18$ displays a similar behaviour to that of $17\_11$ (see Fig. \ref{fig:LG_pressure_profile}), where pressure drops rapidly towards the outskirts of the LG, at a distance beyond the MW's $R_{200}$. In this realisation, however, the two galaxies are further apart ($d = 866\,$kpc in $9\_18$ versus $675\,$kpc in $17\_11$). As a result, pressure decreases more in the region between the galaxies, since their CGMs are not as close to overlapping as in $17\_11$.

\begin{figure*}[h!]
    \centering
    \includegraphics[width=.4\columnwidth]{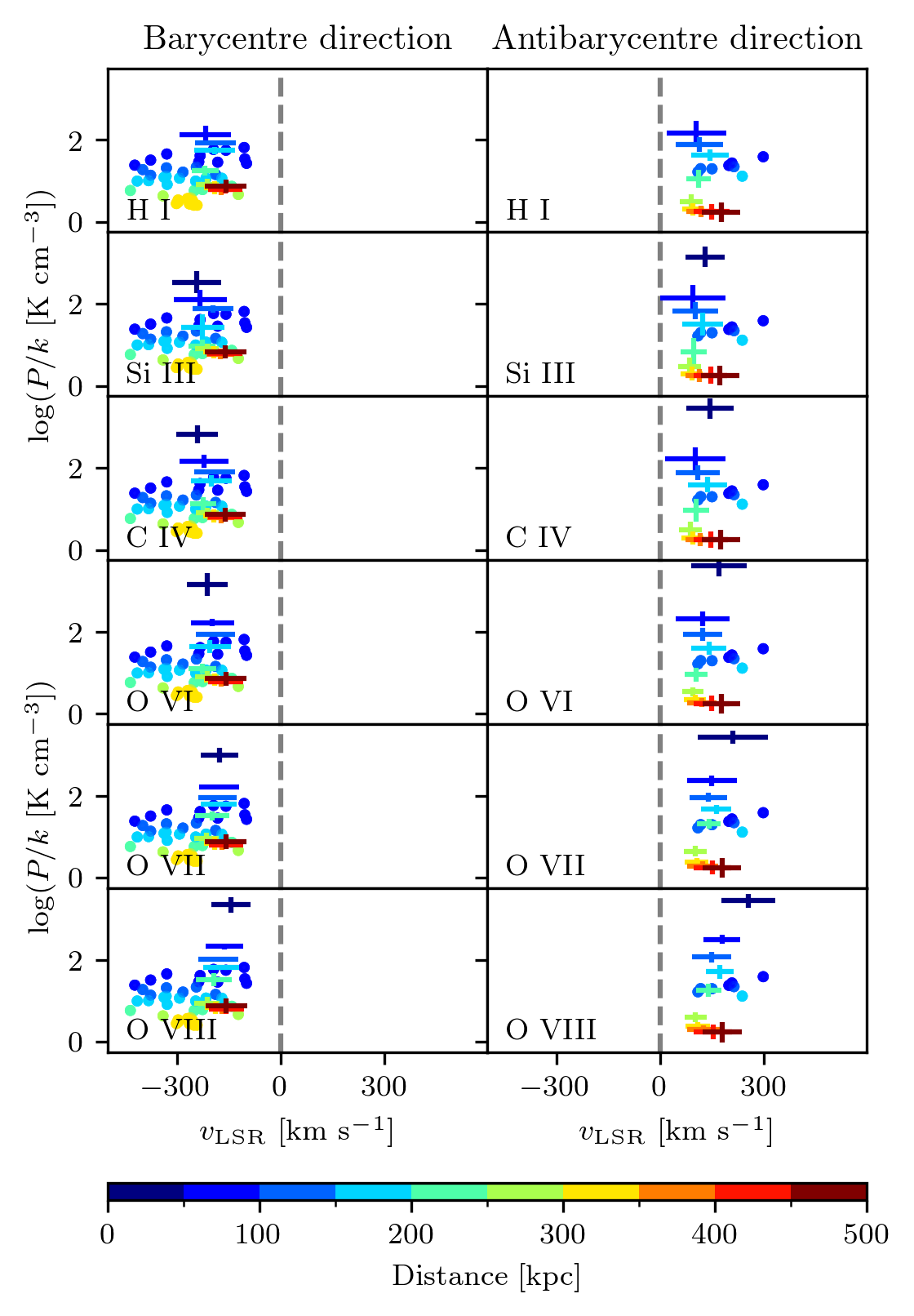}
    \includegraphics[width=.4\columnwidth]{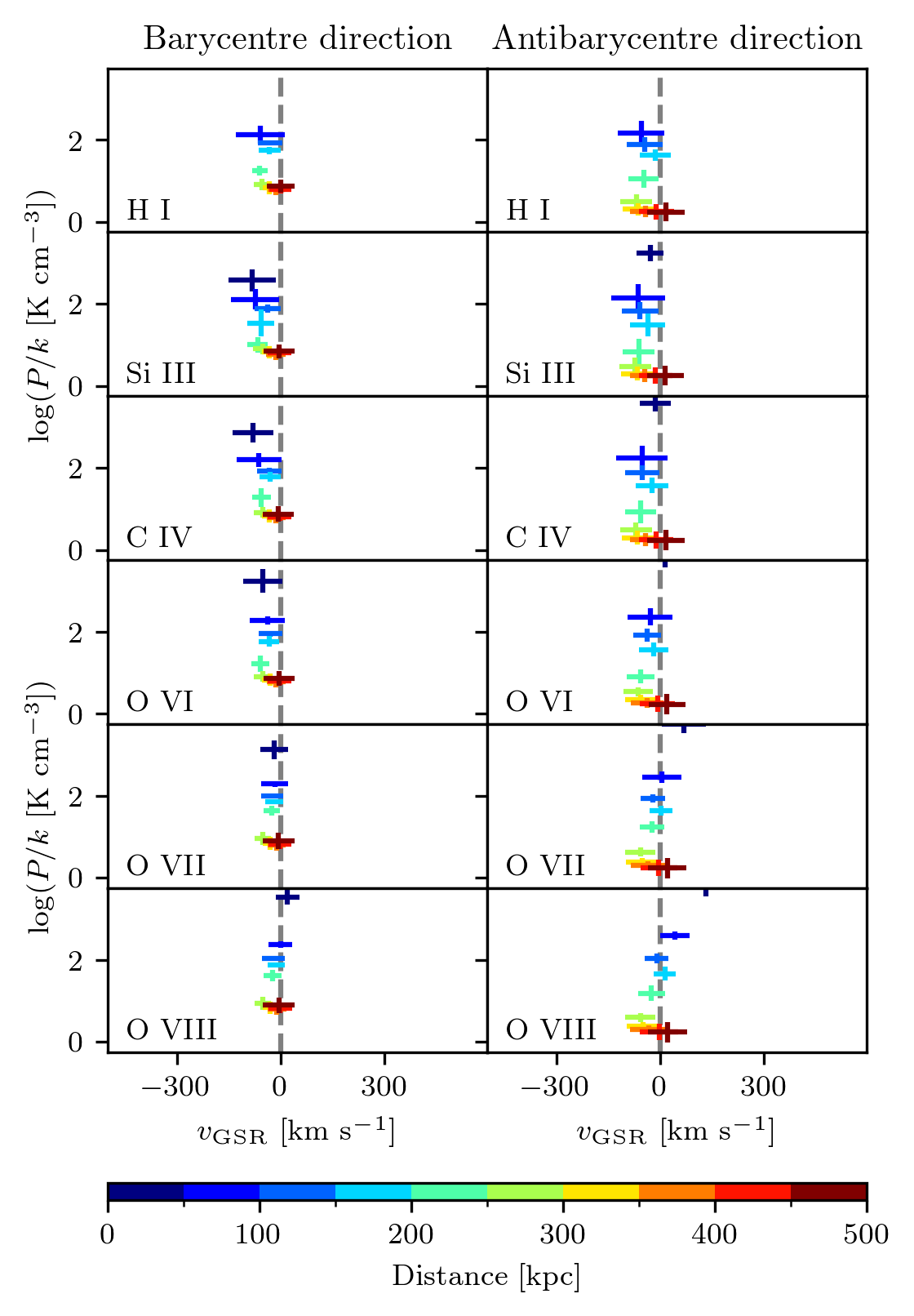}
    \caption{Distance-binned pressure and line-of-sight velocity averages for realisation $9\_18$ in the barycentre and anti-barycentre directions for the LSR (left panel) and the GSR (right panel) reference frames. Overplotted as dots in the left panel are the $v_{\rm LSR}$ and $P/k$ values derived in \cite{Bouma19} for the absorber selection studied in that work.}
    \label{fig:app_pressure_crosses_9_18}
\end{figure*}

In Fig.~\ref{fig:app_pressure_crosses_9_18}, we show the analogous of Fig.~\ref{fig:pressure_crosses} in the case of realisation $9\_18$. In the left panel, ion pressures are compared with the absorber sample of \cite{Bouma19} in the barycentre and anti-barycentre directions, which display a better match to observations than realisation $17\_11$. In the right panel, all line-of-sight velocities are converted to the GSR frame, thus highlighting the relative motion between the MW and M31 candidates. Note that for distances $r\lesssim R_{200}$ cold gas-tracers such as H\,{\sc i}, Si\,{\sc iii}, C\,{\sc iv} are preferentially accreting onto the MW halo. Interestingly, the oxygen lowest ion shows a similar trend, although this signal is washed out for the hot gas-tracing ions O\,{\sc vii} and O\,{\sc viii}. 

\section{Hydrostatic equilibrium in the $17\_11$ MW halo}
\label{app:HSE}

\begin{figure*}[h!]
    \centering
    \includegraphics[scale=1]{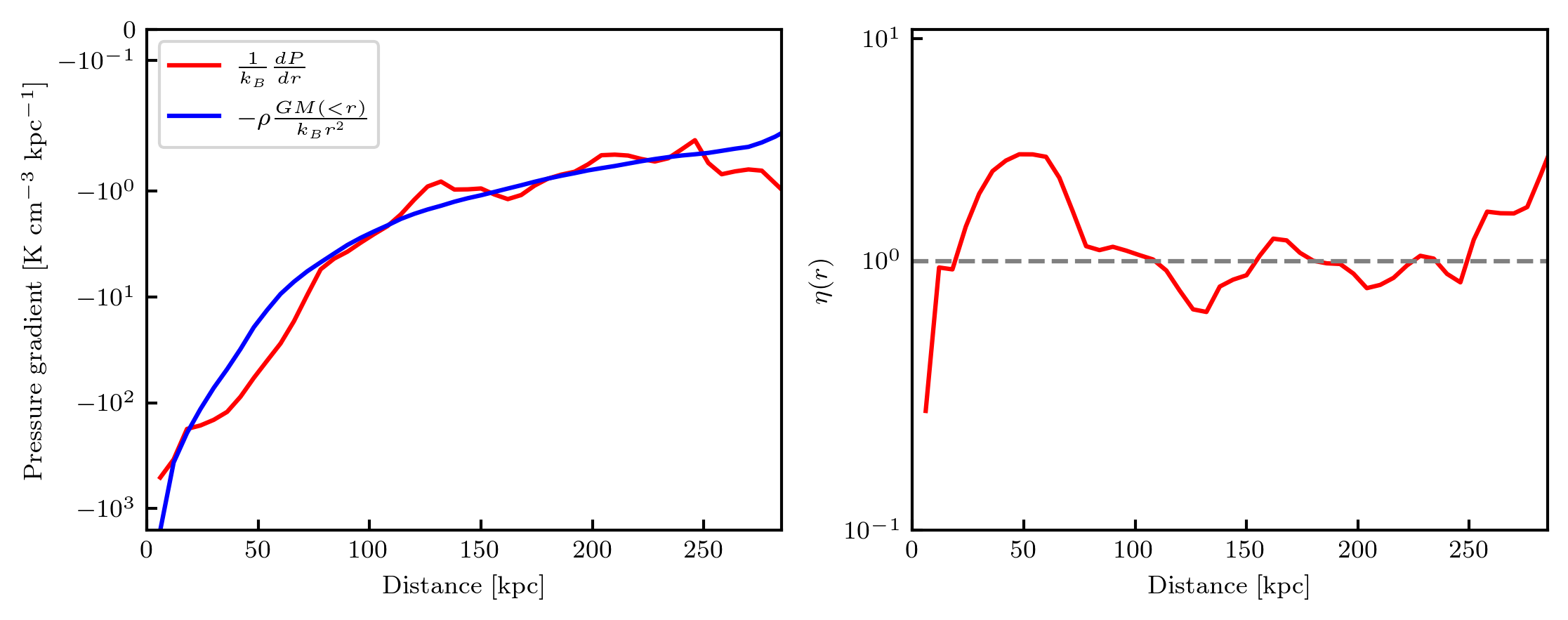}
    \caption{Radial profile of the pressure and gravitational terms in the hydrostatic equilibrium equation (left panel) and their ratio $\eta (r)$ (right panel) in the $17\_11$ MW halo gas at temperatures above $10^6$~K.}
    \label{fig:HSE}
\end{figure*}

To assess the hydrodynamical state of the MW halo gas in the $17\_11$ realisation, we present, in the left panel of Fig. \ref{fig:HSE}, the pressure and gravitational terms of the hydrostatic equilibrium (HSE) equation as a function of radius for hot gas ($T>10^6\,$K) in the halo. Specifically, in spherical coordinates, we can write:

    $$
    \frac{dP}{dr} = -\rho \frac{GM(<r)}{r^2}.
    $$

The deviation from HSE can be quantified by defining the ratio between both sides of the equation above, namely: 

        $$
        \eta(r) \vcentcolon = \bigg|\frac{dP}{dr}\bigg|\left(\rho\frac{GM(<r)}{r^2}\right)^{-1},
        $$

\noindent as it is shown in the right panel of Fig. \ref{fig:HSE}. If $\eta \approx 1$ we expect the halo to be close to HSE. The MW halo in this realisation does not show a significant departure from HSE throughout most of its extent. The exception would be the region at $r\approx50\,$kpc, where $\eta>1$, meaning that pressure gradients should win over gravitational forces, most likely related to the reported outflows in this simulation. The detailed study of the dynamical state of gas in the halo is out of the scope of this paper.\\

\end{appendix}

\end{document}